\documentclass[pra,showpacs,nofootinbib]{revtex4}

\usepackage[latin1]{inputenc}
\usepackage{amsmath,amssymb,amscd,bbm,epsfig}
\usepackage[all,cmtip]{xy}

\newcommand{\bigintlim}{\int}
\newcommand{\bignone}{\,}

\newcommand{\mathd}{\mathrm{d}}
\newcommand{\mathe}{\mathrm{e}}

\newcommand{\tmem}[1]{{\em #1\/}}
\newcommand{\tmmathbf}[1]{\ensuremath{\boldsymbol{#1}}}
\newcommand{\tmop}[1]{\ensuremath{\operatorname{#1}}}
\newcommand{\tmtextbf}[1]{{\bfseries{#1}}}
\newcommand{\tmtextit}[1]{{\itshape{#1}}}

\newcommand{\tmscript}[1]{\scriptstyle{#1}}

\begin{document}

\title{Non-Markovian dynamics for bipartite systems}
\author{Bassano \surname{Vacchini}}
\email{bassano.vacchini@mi.infn.it}
\affiliation{Dipartimento di Fisica dell'Università di Milano and INFN Sezione di Milano, Via Celoria 16, 20133 Milano, Italy}
\date{\today}

\begin{abstract}
  We analyze the appearance of non-Markovian effects in the dynamics of a
  bipartite system coupled to a reservoir, which can be described within a
  class of non-Markovian equations given by a generalized Lindblad structure.
  A novel master equation, which we term quantum Bloch-Boltzmann equation, is
  derived, describing both motional and internal states of a test particle in
  a quantum framework. When due to the preparation of the system or to
  decoherence effects one of the two degrees of freedom is amenable to a
  classical treatment and not resolved in the final measurement, though
  relevant for the interaction with the reservoir, non-Markovian behaviors
  such as stretched exponential or power law decay of coherences can be put
  into evidence.
\end{abstract}
\pacs{03.65.Yz, 05.20.Dd, 03.75.-b, 42.50.Lc}
\maketitle

\section{Introduction\label{sec:intro}}

The complete isolation of quantum mechanical systems, which should arise by a
perfect shielding from the environment, can of course in general only be an
idealization. The study of the dynamics of open quantum systems then naturally
becomes of great interest {\cite{Breuer2007}}, especially when it comes to a
realistic description of experimental situations. While for the case of a
closed quantum system the time evolution is given by a one-parameter unitary
group characterized by a self-adjoint Hamiltonian, the situation is more
involved for an open quantum system, where dynamical evolutions including
irreversible effects like dissipation and decoherence must also be considered.
The possible structures of dynamical equations for an open quantum system are
not known in full generality, despite the huge efforts devoted to the problem.
A well-known result of paramount importance has been obtained for Markovian
dynamics, asking the mapping giving the dynamics to be a completely positive
quantum dynamical semigroup. The expression of the generators of such
semigroups, which gives the master equation for the statistical operator of
the system, has been fully characterized {\cite{Gorini1976a,Lindblad1976a}},
providing a reference structure, often referred to as Lindblad equation. Such
Lindblad type master equations ensure a well-defined time evolution,
preserving in particular the positivity of the statistical operator. The
different terms and operators appearing in it are often naturally amenable to
a direct physical interpretation. Moreover analytical approaches are often
feasible, and when this is not the case numerical studies can always be
performed, by considering Monte Carlo simulations of suitable stochastic
differential equations associated to the master equation via a particular
unraveling.

Such a general and physically transparent characterization is not
available for master equations describing a non-Markovian dynamics.
However systems exhibiting non-Markovian dynamics, such as memory
effects and decay behaviors other than exponential, are also of great
interest both for practical applications and from a conceptual
standpoint. In this spirit major efforts have been devoted to derive
possibly general classes of master equations, which while providing
well-defined time evolutions also describe non-Markovian effects.
Various difficulties appear in this connection. In particular it is
important to provide a link between the operators entering the
structure of generalized master equations and quantities of physical
relevance characterizing the environment and its coupling to the
system. General classes of non-Markovian master equations have been
obtained in the literature
{\cite{Budini2004a,Shabani2005a,Sudarshan2008a}}, also pointing to
possible physical applications. In particular the analysis of the
interaction of a quantum system with a structured reservoir, performed
via a time-convolutionless projection operator technique relying on
the use of correlated projection operators adapted to the structured
reservoir {\cite{Breuer2006a}}, has led to point out a generalized
Lindblad structure {\cite{Breuer2007a}}. This generalized Lindblad
structure describes a non-Markovian dynamics on states which are given
by classical convex mixtures of subcollections, that is positive trace
class operators with trace equal or less than one, naturally appearing
in the description of quantum experiments {\cite{Vacchini2007a}}.
Master equations of this form have already been proposed in an utterly
different context in order to introduce the notion of event in the
description of quantum mechanical systems {\cite{Blanchard1995a}}, for
the purpose of better understanding the interplay between classical
and quantum description of physical reality. More recently and to the
point similar equations have been considered for the statistical
operator of an active atom interacting through collisions with a gas,
when describing in a classical way the centre of mass degrees of
freedom {\cite{Alicki2003a}}.

General physical mechanisms leading to the appearance of such generalized
Lindblad structures which can account for non-Markovian effects have already
been conceived. This is the case if one studies the dynamics of an open system
coupled to a structured reservoir using the abovementioned
time-convolutionless projection operator technique, provided the projectors
used in obtaining the reduced equations of motion do project on classically
correlated states between system and environment, rather than simply on a
factorized state, as in the common wisdom
{\cite{Breuer2006a,Fischer2007a,Breuer2007b}}. Another natural situation
leading to this generalized Lindblad structure appears in what has been called
generalized Born-Markov approximation {\cite{Budini2005a}}. Here one considers
the usual second order perturbation scheme, but once again the state of system
and bath is supposed not to be factorized, but rather given by a convex
mixture of factorized states. The indexes of the mixture are related to the
structure of the bath and of the interaction Hamiltonian. Further work has
traced back the derivation of non-Markovian equations of this form to the
existence of extra unobserved degrees of freedom mediating the entanglement
between the considered system and a Markovian reservoir
{\cite{Budini2005b,Budini2006a}}. Earlier work {\cite{Esposito2003a}} also led
to this kind of non-Markovian master equations for a system interacting with
an environment with a finite heat capacity, so that energy exchanges between
system and reservoir also affect the energy distribution of the reservoir. It
has been recently shown that the same quantum master equation can also be
derived in a physically more transparent manner by means of the projection
superoperator technique {\cite{Esposito2007a}}.

In the present paper we show how such generalized master equations naturally
arise by considering a bipartite quantum system interacting with a Markovian
reservoir, whenever decoherence effects or superselection rules affect only
one kind of degrees of freedom of the bipartite system. A non-Markovian
behavior then appears when the thus emerged classical label is not resolved in
the final measurement. The analysis is done by means of a concrete and
relevant physical example. We consider the dynamics of a quantum test
particle, whose internal and centre of mass degrees of freedom are both
described quantum mechanically, interacting e.g. with a gaseous background.
The appearance of non-Markovian features is related to the involvement of both
internal and centre of mass degrees of freedom in the scattering amplitude
which describes the coupling between bipartite system and environment. As a
first step a novel quantum master equation is heuristically derived, which
extends previous work on the quantum linear Boltzmann equation
{\cite{Vacchini2000a,Vacchini2001b,Hornberger2006b,Hornberger2008a}},
focussing on a quantum description of the centre of mass degrees of freedom,
and on the Bloch-Boltzmann equation
{\cite{Alicki2003a,Dumcke1985a,Snider1998a,Kryszewski2006a,Hornberger2007b}},
which describes in a classical way the motion of the test particle, but
retains a quantum expression for the dynamics of its internal degrees of
freedom. According to this terminology the obtained equation is termed quantum
Bloch-Boltzmann equation. Two limiting situations then naturally appears. When
decoherence affects more strongly or equivalently on a shorter time scale the
centre of mass degrees of freedom, a generalized Lindblad structure
corresponding to the Bloch-Boltzmann equation appears, in which the momentum of
the test particle is treated classically. This equation describes non-Markovian
effects when the final measurement only affects the internal degrees of
freedom. In a similar way, when the experimental effort is devoted to study
quantum superpositions of motional degrees of freedom, e.g. in interferometers
for massive particles studying robustness of their quantum behavior,
non-Markovian features can appear if the internal degrees of freedom influence
the collisional scattering cross section but are later not observed in the
assessment of the interference pattern.

The paper is organized as follows. In Sect. \ref{sec:gl} we describe
mathematical framework and expression of the generalized Lindblad structure,
in Sect. \ref{sec:qbbe} we outline the derivation of the quantum
Bloch-Boltzmann equation, involving both centre of mass and internal degrees
of freedom, starting from the classical linear Boltzmann equation and the
known expression of the quantum linear Boltzmann equation involving only the
motional degrees of freedom. Sect. \ref{sec:nm} is then devoted to consider
the reduced dynamics of either internal or centre of mass degrees of freedom,
for which a generalized Lindblad structure follows, showing by means of
example the appearance of non-Markovian behaviors. Finally in Sect.
\ref{sec:ceo} we briefly comment on our results.

\section{Generalized Lindblad structure\label{sec:gl}}

We now want to introduce the abovementioned non-Markovian generalization of
the Lindblad structure, which is easily obtained considering the standard
theorem of Gorini, Kossakowski, Sudarshan and Lindblad for a special choice of
Hilbert space for the open system and of the expression of its statistical
operator. Let us consider a bipartite quantum system described on a Hilbert
space $\mathcal{H} \otimes \mathcal{H}_B$, with $\mathcal{H}_{}$ and
$\mathcal{H}_B$ separable. Exploiting the isomorphism of $\mathcal{H}_B$ with
either of the possible physically relevant choices of Hilbert space, such as
$\mathbbm{C}^n$, $l^2 \left( \mathbbm{C} \right)$ or $L^2 \left(
\mathbbm{R}^{} \right)$, the tensor product can be expressed as a direct sum
or direct integral, thus naturally introducing a label $\alpha$. For the case
in which $\mathcal{H}_B$ describes a system with $n$ degrees of freedom one
can consider the two equivalent constructions of the same bipartite Hilbert
space
\begin{eqnarray}
  \mathcal{H} \otimes \mathbbm{C}^n & = & \bigoplus^n_{\alpha = 1} \bignone
  \mathcal{H},  \label{eq:cn}
\end{eqnarray}
and similarly for $l^2 \left( \mathbbm{C} \right)$, replacing the finite sum
with a series. On similar grounds a continuous index appears for the case
\begin{eqnarray}
  \mathcal{H} \otimes \text{$L^2 \left( \mathbbm{R}^{} \right)$} & = &
  \int^{\oplus}_{\mathbbm{R}} d \alpha \bignone \mathcal{H}, 
  \label{eq:continuo}
\end{eqnarray}
exploiting the notion of direct integral of Hilbert spaces (see e.g.
{\cite{Alicki2001a}}). It is then of course possible to consider a statistical
operator for the bipartite system, that is to say a positive, trace class
operator on $\mathcal{H} \otimes \mathcal{H}_B$, normalized to one.
Considering for the sake of example the case of $\mathcal{H} \otimes
\mathbbm{C}^n = \bigoplus^n_{\alpha = 1} \bignone \mathcal{H}$, and denoting
by $\varrho$ the statistical operator of the system one can consider the
general expression of the Lindblad equation for $\varrho$. Let us now restrict
however to statistical operators whose matrix representation is block
diagonal, so that they can equivalently be written as \ $\varrho =
\sum^n_{\alpha = 1} \rho_{\alpha} \otimes | \alpha \rangle \langle \alpha |$
or as $\varrho = \left( \rho_1, \ldots, \rho_{\alpha}, \ldots, \rho_n
\right)$. The index $\alpha$ can now really be interpreted as a classical
label indexing the various subcollections $\rho_{\alpha} \in
\mathcal{\mathcal{T \mathcal{C}_{} \left( \mathcal{H} \right)}}$, which are
given by positive, trace class operators on $\mathcal{H}$ with trace less or
equal than one. Such a block diagonal statistical operator
\begin{eqnarray}
  \text{$\varrho$} & = & \left( \rho_1, \ldots, \rho_{\alpha}, \ldots, \rho_n
  \right) \nonumber
\end{eqnarray}
fixed by the set of subcollections $\left\{ \rho_{\alpha} \right\}_{\alpha =
1, \ldots, n}$ is normalized according to
\begin{eqnarray}
  \tmop{Tr} \rho & = & \sum^n_{\alpha = 1} \tmop{Tr}_{\mathcal{\mathcal{H}}}
  \rho_{\alpha} = 1 \text{} .  \label{eq:norm}
\end{eqnarray}
The set of trace class operators which are block diagonal is a closed
subalgebra of the set of all trace class operators, whose dual space is given
by the closed subalgebra of bounded operators also having a block diagonal
structure. Equivalently one can say that this subclass of statistical
operators only provides information on the expectation values of observables
diagonal with respect to the label $\alpha$, which thus correspond to the only
relevant variables for the system under consideration. Considering a
statistical operator in the subalgebra of block diagonal trace class operators
$\rho \in \mathcal{TC}_{\tmop{diag}} \left( \mathcal{H} \otimes \mathbbm{C}^n
\right)$ and an observable given by a block diagonal bounded operator $B \in
\mathcal{B}_{\tmop{diag}} \left( \mathcal{H} \otimes \mathbbm{C}^n \right)$
\begin{eqnarray}
  B & = & \left( \mathsf{B}_1, \ldots, \mathsf{B}_{\alpha}, \ldots,
  \mathsf{B}_n \right), \nonumber
\end{eqnarray}
with $\mathsf{B}_{\alpha} \in \mathcal{B} \left( \mathcal{H} \right)$, the
duality relation is given by
\begin{eqnarray}
  \langle B, \varrho \rangle & = & \sum^n_{\alpha = 1}
  \tmop{Tr}_{\mathcal{\mathcal{H}}} \left( \mathsf{B}_{\alpha} \rho_{\alpha}
  \right) \text{} .  \label{eq:dual}
\end{eqnarray}

It is now of interest to consider the expression of the generator of a quantum
dynamical semigroup acting on this bipartite space when applied to such block
diagonal states or observables, with the further constraint that the time
evolved state or observable still preserves this simple block diagonal
structure, thus defining a dynamics which remains within the spaces
$\mathcal{T} \mathcal{C}_{\tmop{diag}} \left( \mathcal{H} \otimes
\mathbbm{C}^n \right)$ or $\mathcal{B}_{\tmop{diag}} \left( \mathcal{H}
\otimes \mathbbm{C}^n \right)$ respectively. In the Schr\"odinger picture this
generalized Lindblad structure can be written in terms of coupled equations
for the different subcollections $\rho_{\alpha}$ according to
{\cite{Blanchard1995a,Breuer2007a}}
\begin{eqnarray}
  \frac{\mathd}{\tmop{dt}} \rho_{\alpha} & = & - \frac{i}{\hbar} [
  \mathsf{H}^{\alpha}, \rho_{\alpha}] + \sum_{\lambda} \sum_{\beta = 1}^n
  \left[ \mathsf{R}^{\alpha \beta}_{\lambda} \rho_{\beta} \mathsf{R}^{\alpha
  \beta}_{\lambda} \phantom{}^{\dag} - \frac{1}{2} \left\{ \mathsf{R}^{\beta
  \alpha}_{\lambda} \phantom{}^{\dag} \mathsf{R}^{\beta \alpha}_{\lambda},
  \rho_{\alpha} \right\} \right],  \label{eq:S}
\end{eqnarray}
leading due to the duality relation Eq. (\ref{eq:dual}) to the following
equations in Heisenberg picture for the components $\mathsf{B}_{\alpha}$ of a
block diagonal observable
\begin{eqnarray}
  \frac{\mathd}{\tmop{dt}} \mathsf{B}_{\alpha} & = & + \frac{i}{\hbar} [
  \mathsf{H}^{\alpha}, \mathsf{B}_{\alpha}] + \sum_{\lambda} \sum_{\beta =
  1}^n \left[ \mathsf{R}^{\beta \alpha}_{\lambda} \phantom{}^{\dag}
  \mathsf{B}_{\beta} \mathsf{R}^{\beta \alpha}_{\lambda} - \frac{1}{2} \left\{
  \mathsf{R}^{\beta \alpha}_{\lambda} \phantom{}^{\dag} \mathsf{R}^{\beta
  \alpha}_{\lambda}, \mathsf{B}_{\alpha} \right\} \right] .  \label{eq:H}
\end{eqnarray}
In Eqs. (\ref{eq:S}) and (\ref{eq:H}) the index $\alpha$ runs from $1$ to $n$,
the operators $\mathsf{H}^{\alpha}$ are self-adjoint on $\mathcal{H}$ and
$\mathsf{R}^{\beta \alpha}_{\lambda}$ are operators on $\mathcal{H}$, with
$\lambda$ a further index labelling the various Lindblad operators.

Introducing the mapping
\begin{eqnarray}
  \text{$\mathcal{L \varrho}$} & = & \left( \frac{\mathd}{\tmop{dt}} \rho_1,
  \ldots, \frac{\mathd}{\tmop{dt}} \rho_{\alpha}, \ldots,
  \frac{\mathd}{\tmop{dt}} \rho_n \right) \nonumber
\end{eqnarray}
one can therefore write for the time evolution in Schr\"odinger picture
\begin{eqnarray}
  \varrho (t) & = & \left( \rho_1 (t), \ldots, \rho_{\alpha} (t), \ldots,
  \rho_n (t) \right) \nonumber\\
  & = & e^{t \mathcal{\mathcal{L}}} \varrho_{} (0) \nonumber\\
  & = & e^{t \mathcal{\mathcal{L}}} \left( \rho_1 (0), \ldots, \rho_{\alpha}
  (0), \ldots, \rho_n (0) \right), \nonumber
\end{eqnarray}
and similarly for the Heisenberg picture using the mapping $\mathcal{L}'$ dual
to $\mathcal{L}$ according to the relation Eq. (\ref{eq:dual}). Now Eq.
(\ref{eq:S}) provides a Markovian set of equations for the statistical
operator $\varrho (t) = \left( \rho_1 (t), \ldots, \rho_{\alpha} (t), \ldots,
\rho_n (t) \right)$ on $\mathcal{H} \otimes \mathbbm{C}^n$, but a
non-Markovian dynamics for the statistical operator
\begin{eqnarray}
  w (t) & = & \sum^n_{\alpha = 1} \rho_{\alpha} (t),  \label{eq:rhonm}
\end{eqnarray}
which is a statistical operator on the Hilbert space $\mathcal{H}$ only. In
particular it is not possible to define a mapping from $w (0)$ to $w (t)$
according to the non-commutativity of the following diagram
 \begin{displaymath}
 \xymatrix{
   \varrho(0)=\left( \rho_1 (0),   \ldots, \rho_n (0) \right) \ar[d] \ar[r]^{\exp({t \cal{L}})} & \varrho(t)=\left( \rho_1 (t),   \ldots, \rho_n (t) \right) \ar[d] \\
   w(0)=\sum_{\alpha=1}^n \rho_\alpha(0) \ar[r]|\not & w(t)=\sum_{\alpha=1}^n \rho_\alpha(t)}
 \end{displaymath}
arising because of the loss of information in going from $\varrho (t)$ to $w
(t)$. The set of equations given by Eq. (\ref{eq:S}) thus provides a
non-Markovian dynamics for the statistical operator $w (t)$ supposed to be
expressible at any time as a mixture of subcollections $\rho_{\alpha} (t)$, or
equivalently as a convex combination with weights $p_{\alpha} (t) =
\tmop{Tr}_{\mathcal{H}} \rho_{\alpha} (t)$ of statistical operators given by
$w_{\alpha} (t) = \rho_{\alpha} (t) / \tmop{Tr}_{\mathcal{H}} \rho_{\alpha}
(t)$. This last standpoint stresses the appearance of the classical
probability distribution $\left\{ p_{\alpha} (t) \right\}_{\alpha = 1, \ldots,
n}$, which justifies the name random Lindblad equations or Lindblad rate
equations {\cite{Budini2005a,Budini2006a}}, also given to equations falling
within the class given by Eq. (\ref{eq:S}).

The statistical operator $w (t)$ can arise in a twofold way. Either by taking
the trace of a block diagonal $\varrho (t)$ with respect to $\mathbbm{C}^n$,
corresponding to a situation in which one considers a reduced dynamics of the
bipartite system with respect to the degrees of freedom which behave
effectively in a classical way, or by assuming that the state of the system
under study living in the Hilbert space $\mathcal{H}$ is specified at the
initial time as a convex combination of $n$ statistical operators with
suitable weights, and retains this form throughout the dynamics. The first
type of realization makes it intuitively clear why Eq. (\ref{eq:S})
encompasses non-Markovian situations. By looking at the time evolution of $w
(t)$ only one is considering a restricted set of variables with respect to the
full collection $\{\rho_{\alpha} \left( t \right) \}_{\alpha = 1, \ldots, n}$,
for which the time evolution law would be Markovian. The set of relevant
physical variables then determines whether or not the dynamics is Markovian.
Statistical operators of the form given by Eq. (\ref{eq:rhonm}) naturally
appear in connection with a structured reservoir, the label $\alpha$ being
then connected to a characterization of the reservoir itself, e.g. labelling
different energy bands. This is the case both when considering a projection
operator technique assuming classical correlated states between system and
reservoir {\cite{Breuer2006a}}, or more simply in the so called generalized
Born-Markov approximation {\cite{Budini2005a}}, using a classically correlated
state in the derivation of the master equation to second order in the
perturbation. As we shall argue below statistical operators of this form also
appear when considering a bipartite system interacting with a reservoir, when
due to decoherence one of the two kind of degrees of freedom behaves
classically, and despite characterizing the initial preparation and being
relevant for the interaction with the environment, cannot later be resolved by
the measurement apparatus. Needless to say the formal scheme developed above
can also be implemented when the label $\alpha$ runs over a countable set or
even in the continuum, corresponding to a Hilbert space construction as
depicted in Eq. (\ref{eq:continuo}).

\section{Derivation of quantum Bloch-Boltzmann equation\label{sec:qbbe}}

We now address the issue of the derivation of a master equation
describing the dynamics of both internal and centre of mass degrees of
freedom of a quantum test particle interacting through collisions with
a reservoir such as a background gas. Here we will outline a heuristic
derivation, leaving a more detailed and microscopic one for a later
publication, in particular we will concentrate only on the incoherent
terms corresponding to gain and loss term in the classical case,
leaving aside the Hamiltonian contributions due to kinetic term and
forward scattering. To do this we will build on the known results for
the quantum linear Boltzmann equation \
{\cite{Vacchini2000a,Vacchini2001b,Hornberger2006b,Hornberger2008a}},
describing in a quantum framework the centre of mass degrees of
freedom, and on the so called Bloch-Boltzmann equation
{\cite{Alicki2003a,Dumcke1985a,Snider1998a,Kryszewski2006a,Hornberger2007b}},
which accounts for a semiclassical description of both internal and
centre of mass degrees of freedom.  As a starting point we take the
equations of motion considered in \cite{Alicki2003a} for the
collection $\left\{ \text{$\rho \left( \tmmathbf{P} \right)$}
\right\}_{\tmmathbf{P} \in \mathbbm{R}^3}$ of trace class operators on
the space $\mathbbm{C}^n$ of internal degrees of freedom, labelled
with the continuous index $\tmmathbf{P}$ now characterizing the
classical momentum of the test particle. This result extends the work
in \cite{Dumcke1985a,Hornberger2007b} by considering also the dynamics
of the centre of mass degrees of freedom of the test particle, which
in \cite{Dumcke1985a,Hornberger2007b} is supposed to have infinite
mass and therefore to be at rest.
It is most convenient to write the equation putting into
evidence the constraints due to momentum and energy conservation in
the single interaction events, thus obtaining
\begin{eqnarray}
  \frac{\mathd}{\tmop{dt}} \rho (\tmmathbf{P}) & = &
  \frac{n_{\tmop{gas}}}{m^2_{\ast}} \sum_{\tmscript{\begin{subarray}{c}
    ijkl\\
    \mathcal{E}_{ij} = \mathcal{E}_{kl}
  \end{subarray}}} \bigintlim \mathd \tmmathbf{P}' \int \mathd \tmmathbf{p}' \int
  \bignone \mathd \tmmathbf{p} \hspace{0.2em} \mu_{\beta} \left(
  \tmmathbf{p}'_{} \right)  \nonumber\\
  &  & \times \delta \left( \frac{P'^2}{2 M} + \frac{p'^2}{2 m} + \hbar
  \omega_j - \frac{P^2}{2 M} - \frac{p^2}{2 m} - \hbar \omega_i \right)
  \delta^3 \left( \tmmathbf{P}' +\tmmathbf{p}' -\tmmathbf{P}-\tmmathbf{p}
  \right) \nonumber\\
  &  & \times f_{ij} \left( \tmop{rel} \left( \tmmathbf{p}, \tmmathbf{P}
  \right), \tmop{rel} \left( \tmmathbf{p}', \tmmathbf{P}' \right) \right)
  f^{\ast}_{kl} \left( \tmop{rel} \left( \tmmathbf{p}, \tmmathbf{P} \right),
  \tmop{rel} \left( \tmmathbf{p}', \tmmathbf{P}' \right) \right) E_{ij} \rho
  (\tmmathbf{P}') E^{^{\dag}}_{kl} \nonumber\\
  &  & - \frac{1}{2} \frac{n_{\tmop{gas}}}{m^2_{\ast}}
  \sum_{\tmscript{\begin{subarray}{c}
    ijkl\\
    \mathcal{E}_{ij} = \mathcal{E}_{kl}
  \end{subarray}}} \bigintlim \mathd \tmmathbf{P}' \int \mathd \tmmathbf{p}' \int
  \bignone \mathd \tmmathbf{p} \hspace{0.2em} \mu_{\beta} \left(
  \tmmathbf{p}_{} \right)  \nonumber\\
  &  & \times \delta \left( \frac{P^2}{2 M} + \frac{p^2}{2 m} + \hbar
  \omega_j - \frac{P'^2}{2 M} - \frac{p'^2}{2 m} - \hbar \omega_i \right)
  \delta^3 \left( \tmmathbf{P}+\tmmathbf{p}-\tmmathbf{P}' -\tmmathbf{p}'
  \right) \nonumber\\
  &  & \times f_{ij} \left( \tmop{rel} \left( \tmmathbf{p}', \tmmathbf{P}'
  \right), \tmop{rel} \left( \tmmathbf{p}, \tmmathbf{P} \right) \right)
  f^{\ast}_{kl} \left( \tmop{rel} \left( \tmmathbf{p}', \tmmathbf{P}' \right),
  \tmop{rel} \left( \tmmathbf{p}, \tmmathbf{P} \right) \right) \left\{
  E^{^{\dag}}_{kl} E_{ij}, \rho (\tmmathbf{P}) \right\} .  \label{eq:clbeMB}
\end{eqnarray}
In this equation $M$ and $m$ denote the mass of test particle and gas
particles respectively, $m^{}_{\ast}$ is the reduced mass, $\mu_{\beta} \left(
\tmmathbf{p}_{} \right)$ is the Maxwell-Boltzmann distribution of the gas
\begin{eqnarray}
  \mu_{\beta} \left( \tmmathbf{p} \right) & = & \frac{1}{\pi^{3 / 2}
  p_{\beta}^3} \exp \left( - \frac{\tmmathbf{p}^2}{p_{\beta}^2} \right), 
  \label{eq:muMB}
\end{eqnarray}
with $p_{\beta}^{} = \sqrt{2 m / \beta}$ the most probable momentum at
temperature $T = 1 / \left( k_{\text{B}} \beta \right)$, relative momenta are
denoted as
\begin{eqnarray}
  \tmop{rel} \left( \tmmathbf{p}, \tmmathbf{P} \right) & \equiv &
  \frac{m_{\ast}}{m} \tmmathbf{p}- \frac{m_{\ast}}{M} \tmmathbf{P}, 
  \label{eq:reldef}
\end{eqnarray}
and
\begin{eqnarray}
  f_{ij} \left( \tmmathbf{p}, \tmmathbf{p}' \right) & \equiv & f \left(
  \tmmathbf{p}', j \rightarrow \tmmathbf{p}, i \right)  \label{eq:ampl}
\end{eqnarray}
indicates the complex scattering amplitude for a transition from an
{\tmem{in}} state with labels $\tmmathbf{p}', j$ to an {\tmem{out}} state with
labels $\tmmathbf{p}, i$. Finally the matrices $E_{ij}$ correspond to the
mappings between energy eigenstates, providing a basis of operators in
$\mathbbm{C}^n$, according to
\begin{eqnarray}
  E_{ij} & = & |i \rangle \langle j|,  \label{eq:base}
\end{eqnarray}
while $\hbar \omega_j$ is the energy of the $j$-level and $\mathcal{E}_{ij} =
\hbar \omega_i - \hbar \omega_j$ denote the possible transition energies.
Using the delta of momentum conservation Eq. (\ref{eq:clbeMB}) can also be
expressed using as a variable the momentum transfer
$\tmmathbf{Q}=\tmmathbf{P}' -\tmmathbf{P}$ in the single collisions, thus
coming to
\begin{eqnarray}
  \frac{\mathd}{\tmop{dt}} \rho (\tmmathbf{P}) & = &
  \frac{n_{\tmop{gas}}}{m^2_{\ast}} \sum_{\tmscript{\begin{subarray}{c}
    ijkl\\
    \mathcal{E}_{ij} = \mathcal{E}_{kl}
  \end{subarray}}} \bigintlim \mathd \tmmathbf{Q} \int \bignone \mathd
  \tmmathbf{p} \hspace{0.2em} \mu_{\beta} \left( \tmmathbf{p}-\tmmathbf{Q}
  \right) \delta \left( \frac{\left( \tmmathbf{P}+\tmmathbf{Q} \right)^2}{2 M}
  + \frac{(\tmmathbf{p}-\tmmathbf{Q})^2}{2 m} - \frac{P^2}{2 M} - \frac{p^2}{2
  m} + \mathcal{E}_{ij} \right) \nonumber\\
  &  & \times f_{ij} \left( \tmop{rel} \left( \tmmathbf{p}, \tmmathbf{P}
  \right), \tmop{rel} \left( \tmmathbf{p}, \tmmathbf{P} \right) -\tmmathbf{Q}
  \right) f^{\ast}_{kl} \left( \tmop{rel} \left( \tmmathbf{p}, \tmmathbf{P}
  \right), \tmop{rel} \left( \tmmathbf{p}, \tmmathbf{P} \right) -\tmmathbf{Q}
  \right) E_{ij} \rho (\tmmathbf{P}+\tmmathbf{Q}) E^{^{\dag}}_{kl} \nonumber\\
  &  & - \frac{1}{2} \frac{n_{\tmop{gas}}}{m^2_{\ast}}
  \sum_{\tmscript{\begin{subarray}{c}
    ijkl\\
    \mathcal{E}_{ij} = \mathcal{E}_{kl}
  \end{subarray}}} \bigintlim \mathd \tmmathbf{Q} \int \bignone \mathd
  \tmmathbf{p} \hspace{0.2em} \mu_{\beta} \left( \tmmathbf{p} \right) \delta
  \left( \frac{\left( \tmmathbf{P}+\tmmathbf{Q} \right)^2}{2 M} +
  \frac{(\tmmathbf{p}-\tmmathbf{Q})^2}{2 m} - \frac{P^2}{2 M} - \frac{p^2}{2
  m} + \mathcal{E}_{ij} \right) \nonumber\\
  &  & \times f_{ij} \left( \tmop{rel} \left( \tmmathbf{p}, \tmmathbf{P}
  \right) -\tmmathbf{Q}, \tmop{rel} \left( \tmmathbf{p}, \tmmathbf{P} \right)
  \right) f^{\ast}_{kl} \left( \tmop{rel} \left( \tmmathbf{p}, \tmmathbf{P}
  \right) -\tmmathbf{Q}, \tmop{rel} \left( \tmmathbf{p}, \tmmathbf{P} \right)
  \right) \left\{ E^{^{\dag}}_{kl} E_{ij}, \rho (\tmmathbf{P}) \right\} .
  \nonumber
\end{eqnarray}
Introducing now by the notation $\|\tmmathbf{Q}$ and $\perp \tmmathbf{Q}$ the
component of a vector $\tmop{parallel}$ and perpendicular to the momentum
transfer $\tmmathbf{Q}$, so that $\tmmathbf{P}_{\|\tmmathbf{Q}} = \left(
\tmmathbf{P} \cdot \tmmathbf{Q} \right) \tmmathbf{Q}/ Q^2$ and
$\tmmathbf{P}_{\bot \tmmathbf{Q}} =\tmmathbf{P}-\tmmathbf{P}_{\|\tmmathbf{Q}}$
respectively, one has using Eq. (\ref{eq:reldef}) in the delta of energy
conservation
\begin{eqnarray}
  \delta \left( \frac{\left( \tmmathbf{P}+\tmmathbf{Q} \right)^2}{2 M} +
  \frac{(\tmmathbf{p}-\tmmathbf{Q})^2}{2 m} - \frac{P^2}{2 M} - \frac{p^2}{2
  m} + \mathcal{E}_{ij} \right) = \delta \left( \frac{Q^2}{2 m_{\ast}} -
  \frac{\tmmathbf{Q}}{m_{\ast}} \mathbf{\tmmathbf{}} \cdot \tmop{rel} \left(
  \mathbf{\tmmathbf{p}}_{\| \tmmathbf{Q}}, \tmmathbf{P}_{\| \tmmathbf{Q}}
  \right) + \mathcal{E}_{ij} \right),  \label{eq:ec} &  & 
\end{eqnarray}
so that in the integral one can use the replacement
\begin{eqnarray}
  \tmop{rel} \left( \mathbf{\tmmathbf{p}}_{\| \tmmathbf{Q}}, \tmmathbf{P}_{\|
  \tmmathbf{Q}} \right) & = & \frac{1}{2} \left( 1 +
  \frac{\mathcal{E}_{ji}}{Q^2 / \left( 2 m_{\ast} \right)} \right)
  \tmmathbf{Q},  \label{eq:delta}
\end{eqnarray}
and therefore
 \begin{eqnarray}
   \frac{\mathd}{\tmop{dt}} \rho (\tmmathbf{P}) & = &
   \frac{n_{\tmop{gas}}}{m^2_{\ast}} \sum_{\tmscript{\begin{subarray}{c}
     ijkl\\
     \mathcal{E}_{ij} = \mathcal{E}_{kl}
   \end{subarray}}} \bigintlim \mathd \tmmathbf{Q} \int \bignone \mathd
   \tmmathbf{p} \hspace{0.2em} \mu_{\beta} \left( \tmmathbf{p}-\tmmathbf{Q}
   \right) \delta \left( \frac{Q^2}{2 m_{\ast}} - \frac{\tmmathbf{Q}}{m_{\ast}}
   \cdot \tmop{rel} \left( \mathbf{\tmmathbf{p}}_{\| \tmmathbf{Q}},
   \tmmathbf{P}_{\| \tmmathbf{Q}} \right) + \mathcal{E}_{ij} \right)
   \nonumber\\
   &  & \times f_{ij} \left( \tmop{rel} \left( \mathbf{\tmmathbf{p}}_{\perp
   \tmmathbf{Q}}, \tmmathbf{P}_{\perp \tmmathbf{Q}} \right) +
   \frac{\tmmathbf{Q}}{2} + \frac{\mathcal{E}_{ji}}{Q^2 / m_{\ast}}
   \tmmathbf{Q}, \tmop{rel} \left( \mathbf{\tmmathbf{p}}_{\perp \tmmathbf{Q}},
   \tmmathbf{P}_{\perp \tmmathbf{Q}} \right) - \frac{\tmmathbf{Q}}{2} +
   \frac{\mathcal{E}_{ji}}{Q^2 / m_{\ast}} \tmmathbf{Q} \right) \nonumber\\
   &  & \times f^{\ast}_{kl} \left( \tmop{rel} \left(
   \mathbf{\tmmathbf{p}}_{\perp \tmmathbf{Q}}, \tmmathbf{P}_{\perp
   \tmmathbf{Q}} \right) + \frac{\tmmathbf{Q}}{2} + \frac{\mathcal{E}_{lk}}{Q^2
   / m_{\ast}} \tmmathbf{Q}, \tmop{rel} \left( \mathbf{\tmmathbf{p}}_{\perp
   \tmmathbf{Q}}, \tmmathbf{P}_{\perp \tmmathbf{Q}} \right) -
   \frac{\tmmathbf{Q}}{2} + \frac{\mathcal{E}_{lk}}{Q^2 / m_{\ast}}
   \tmmathbf{Q} \right) \nonumber\\
   &  & \times E_{ij} \rho (\tmmathbf{P}+\tmmathbf{Q}) E^{^{\dag}}_{kl}
   \nonumber\\
   &  & - \frac{1}{2} \frac{n_{\tmop{gas}}}{m^2_{\ast}}
   \sum_{\tmscript{\begin{subarray}{c}
     ijkl\\
     \mathcal{E}_{ij} = \mathcal{E}_{kl}
   \end{subarray}}} \bigintlim \mathd \tmmathbf{Q} \int \bignone \mathd
   \tmmathbf{p} \hspace{0.2em} \mu_{\beta} \left( \tmmathbf{p} \right) \delta
   \left( \frac{Q^2}{2 m_{\ast}} - \frac{\tmmathbf{Q}}{m_{\ast}} \cdot
   \tmop{rel} \left( \mathbf{\tmmathbf{p}}_{\| \tmmathbf{Q}}, \tmmathbf{P}_{\|
   \tmmathbf{Q}} \right) + \mathcal{E}_{ij} \right) \nonumber\\
   &  & \times f_{ij} \left( \tmop{rel} \left( \mathbf{\tmmathbf{p}}_{\perp
   \tmmathbf{Q}}, \tmmathbf{P}_{\perp \tmmathbf{Q}} \right) -
   \frac{\tmmathbf{Q}}{2} + \frac{\mathcal{E}_{ij}}{Q^2 / m_{\ast}}
   \tmmathbf{Q}, \tmop{rel} \left( \mathbf{\tmmathbf{p}}_{\perp \tmmathbf{Q}},
   \tmmathbf{P}_{\perp \tmmathbf{Q}} \right) + \frac{\tmmathbf{Q}}{2} +
   \frac{\mathcal{E}_{ij}}{Q^2 / m_{\ast}} \tmmathbf{Q} \right) \nonumber\\
   &  & \times f^{\ast}_{kl} \left( \tmop{rel} \left(
   \mathbf{\tmmathbf{p}}_{\perp \tmmathbf{Q}}, \tmmathbf{P}_{\perp
   \tmmathbf{Q}} \right) - \frac{\tmmathbf{Q}}{2} + \frac{\mathcal{E}_{kl}}{Q^2
   / m_{\ast}} \tmmathbf{Q}, \tmop{rel} \left( \mathbf{\tmmathbf{p}}_{\perp
   \tmmathbf{Q}}, \tmmathbf{P}_{\perp \tmmathbf{Q}} \right) +
   \frac{\tmmathbf{Q}}{2} + \frac{\mathcal{E}_{kl}}{Q^2 / m_{\ast}}
   \tmmathbf{Q} \right) \nonumber\\
   &  & \times \left\{ E^{^{\dag}}_{kl} E_{ij}, \rho (\tmmathbf{P}) \right\} .
   \nonumber
 \end{eqnarray}
One can now perform the translation $_{} \tmmathbf{p} \rightarrow
\tmmathbf{p}+ m\tmmathbf{Q}/ \left( 2 m_{\ast} \right) +
m\tmmathbf{P}_{\|\tmmathbf{Q}} / M + m_{}  \mathcal{E}_{ji} \tmmathbf{Q}/ Q^2$
in the gain term, and similarly for the loss one, which does not affect the
argument of the scattering amplitudes, thus obtaining
\begin{eqnarray}
  \frac{\mathd}{\tmop{dt}} \rho (\tmmathbf{P}) & = & \frac{n_{\tmop{gas}}
  m}{m^2_{\ast}} \sum_{\tmscript{\begin{subarray}{c}
    ijkl\\
    \mathcal{E}_{ij} = \mathcal{E}_{kl}
  \end{subarray}}} \bigintlim \mathd \tmmathbf{Q} \int \bignone \mathd
  \tmmathbf{p} \hspace{0.2em} \mu_{\beta} \left( \tmmathbf{p}+
  \frac{m}{m_{\ast}} \frac{\tmmathbf{Q}}{2} + \frac{m}{M} \left(
  \tmmathbf{P}_{\|\tmmathbf{Q}} -\tmmathbf{Q} \right) +
  \frac{\mathcal{E}_{ij}}{Q^2 / m} \tmmathbf{Q} \right) \delta \left(
  \mathbf{\tmmathbf{Q}} \cdot \tmmathbf{p} \right) \nonumber\\
  &  & \times f_{ij} \left( \tmop{rel} \left( \mathbf{\tmmathbf{p}}_{\perp
  \tmmathbf{Q}}, \tmmathbf{P}_{\perp \tmmathbf{Q}} \right) -
  \frac{\tmmathbf{Q}}{2} + \frac{\mathcal{E}_{ij}}{Q^2 / m_{\ast}}
  \tmmathbf{Q}, \tmop{rel} \left( \mathbf{\tmmathbf{p}}_{\perp \tmmathbf{Q}},
  \tmmathbf{P}_{\perp \tmmathbf{Q}} \right) + \frac{\tmmathbf{Q}}{2} +
  \frac{\mathcal{E}_{ij}}{Q^2 / m_{\ast}} \tmmathbf{Q} \right) \nonumber\\
  &  & \times f^{\ast}_{kl} \left( \tmop{rel} \left(
  \mathbf{\tmmathbf{p}}_{\perp \tmmathbf{Q}}, \tmmathbf{P}_{\perp
  \tmmathbf{Q}} \right) - \frac{\tmmathbf{Q}}{2} + \frac{\mathcal{E}_{kl}}{Q^2
  / m_{\ast}} \tmmathbf{Q}, \tmop{rel} \left( \mathbf{\tmmathbf{p}}_{\perp
  \tmmathbf{Q}}, \tmmathbf{P}_{\perp \tmmathbf{Q}} \right) +
  \frac{\tmmathbf{Q}}{2} + \frac{\mathcal{E}_{kl}}{Q^2 / m_{\ast}}
  \tmmathbf{Q} \right) \nonumber\\
  &  & \times E_{ij} \rho (\tmmathbf{P}-\tmmathbf{Q}) E^{^{\dag}}_{kl}
  \nonumber\\
  &  & - \frac{1}{2} \frac{n_{\tmop{gas}} m}{m^2_{\ast}}
  \sum_{\tmscript{\begin{subarray}{c}
    ijkl\\
    \mathcal{E}_{ij} = \mathcal{E}_{kl}
  \end{subarray}}} \bigintlim \mathd \tmmathbf{Q} \int \bignone \mathd
  \tmmathbf{p} \hspace{0.2em} \mu_{\beta} \left( \tmmathbf{p}+
  \frac{m}{m_{\ast}} \frac{\tmmathbf{Q}}{2} + \frac{m}{M}
  \tmmathbf{P}_{\|\tmmathbf{Q}} + \frac{\mathcal{E}_{ij}}{Q^2 / m}
  \tmmathbf{Q} \right) \delta \left( \mathbf{\tmmathbf{Q}} \cdot \tmmathbf{p}
  \right) \nonumber\\
  &  & \times f_{ij} \left( \tmop{rel} \left( \mathbf{\tmmathbf{p}}_{\perp
  \tmmathbf{Q}}, \tmmathbf{P}_{\perp \tmmathbf{Q}} \right) -
  \frac{\tmmathbf{Q}}{2} + \frac{\mathcal{E}_{ij}}{Q^2 / m_{\ast}}
  \tmmathbf{Q}, \tmop{rel} \left( \mathbf{\tmmathbf{p}}_{\perp \tmmathbf{Q}},
  \tmmathbf{P}_{\perp \tmmathbf{Q}} \right) + \frac{\tmmathbf{Q}}{2} +
  \frac{\mathcal{E}_{ij}}{Q^2 / m_{\ast}} \tmmathbf{Q} \right) \nonumber\\
  &  & \times f^{\ast}_{kl} \left( \tmop{rel} \left(
  \mathbf{\tmmathbf{p}}_{\perp \tmmathbf{Q}}, \tmmathbf{P}_{\perp
  \tmmathbf{Q}} \right) - \frac{\tmmathbf{Q}}{2} + \frac{\mathcal{E}_{kl}}{Q^2
  / m_{\ast}} \tmmathbf{Q}, \tmop{rel} \left( \mathbf{\tmmathbf{p}}_{\perp
  \tmmathbf{Q}}, \tmmathbf{P}_{\perp \tmmathbf{Q}} \right) +
  \frac{\tmmathbf{Q}}{2} + \frac{\mathcal{E}_{kl}}{Q^2 / m_{\ast}}
  \tmmathbf{Q} \right) \nonumber\\
  &  & \times \left\{ E^{^{\dag}}_{kl} E_{ij}, \rho (\tmmathbf{P}) \right\},
  \nonumber
\end{eqnarray}
where we also performed the change of variables $\tmmathbf{Q} \rightarrow
-\tmmathbf{Q}$ in the gain term and used the simple relation $m / \left( 2
m_{\ast} \right) - m / M = 1 - m / \left( 2 m_{\ast} \right)$. Noting that
\begin{eqnarray}
  \int \bignone \mathd \tmmathbf{p}g \left( \tmmathbf{p} \right) \delta \left(
  \mathbf{\tmmathbf{Q}} \cdot \tmmathbf{p} \right) & = & \frac{1}{Q}
  \int_{\tmmathbf{Q}^{\bot}} \mathd \tmmathbf{p}g \left( \tmmathbf{p}_{\bot
  \tmmathbf{Q}} \right),  \label{eq:perp}
\end{eqnarray}
where the integration on the r.h.s. is restricted to momenta of the gas
particle perpendicular to the momentum transfer we obtain the equation
\begin{eqnarray}
  \frac{\mathd}{\tmop{dt}} \rho (\tmmathbf{P}) & = & \frac{n_{\tmop{gas}}
  m}{m^2_{\ast}} \sum_{\tmscript{\begin{subarray}{c}
    ijkl\\
    \mathcal{E}_{ij} = \mathcal{E}_{kl}
  \end{subarray}}} \bigintlim \frac{\mathd \tmmathbf{Q}}{Q}
  \int_{\tmmathbf{Q}^{\bot}} \bignone \mathd \tmmathbf{p} \hspace{0.2em}
  \mu_{\beta} \left( \tmmathbf{p}_{\bot \tmmathbf{Q}} + \frac{m}{m_{\ast}}
  \frac{\tmmathbf{Q}}{2} + \frac{m}{M} \left( \tmmathbf{P}_{\|\tmmathbf{Q}}
  -\tmmathbf{Q} \right) + \frac{\mathcal{E}_{ij}}{Q^2 / m} \tmmathbf{Q}
  \right)  \nonumber\\
  &  & \times f_{ij} \left( \tmop{rel} \left( \mathbf{\tmmathbf{p}}_{\perp
  \tmmathbf{Q}}, \tmmathbf{P}_{\perp \tmmathbf{Q}} \right) -
  \frac{\tmmathbf{Q}}{2} + \frac{\mathcal{E}_{ij}}{Q^2 / m_{\ast}}
  \tmmathbf{Q}, \tmop{rel} \left( \mathbf{\tmmathbf{p}}_{\perp \tmmathbf{Q}},
  \tmmathbf{P}_{\perp \tmmathbf{Q}} \right) + \frac{\tmmathbf{Q}}{2} +
  \frac{\mathcal{E}_{ij}}{Q^2 / m_{\ast}} \tmmathbf{Q} \right) \nonumber\\
  &  & \times f^{\ast}_{kl} \left( \tmop{rel} \left(
  \mathbf{\tmmathbf{p}}_{\perp \tmmathbf{Q}}, \tmmathbf{P}_{\perp
  \tmmathbf{Q}} \right) - \frac{\tmmathbf{Q}}{2} + \frac{\mathcal{E}_{kl}}{Q^2
  / m_{\ast}} \tmmathbf{Q}, \tmop{rel} \left( \mathbf{\tmmathbf{p}}_{\perp
  \tmmathbf{Q}}, \tmmathbf{P}_{\perp \tmmathbf{Q}} \right) +
  \frac{\tmmathbf{Q}}{2} + \frac{\mathcal{E}_{kl}}{Q^2 / m_{\ast}}
  \tmmathbf{Q} \right) \nonumber\\
  &  & \times E_{ij} \rho (\tmmathbf{P}-\tmmathbf{Q}) E^{^{\dag}}_{kl}
  \nonumber\\
  &  & - \frac{1}{2} \frac{n_{\tmop{gas}} m}{m^2_{\ast}}
  \sum_{\tmscript{\begin{subarray}{c}
    ijkl\\
    \mathcal{E}_{ij} = \mathcal{E}_{kl}
  \end{subarray}}} \bigintlim \mathd \tmmathbf{Q} \int \bignone \mathd
  \tmmathbf{p} \hspace{0.2em} \mu_{\beta} \left( \tmmathbf{p}_{\bot
  \tmmathbf{Q}} + \frac{m}{m_{\ast}} \frac{\tmmathbf{Q}}{2} + \frac{m}{M}
  \tmmathbf{P}_{\|\tmmathbf{Q}} + \frac{\mathcal{E}_{ij}}{Q^2 / m}
  \tmmathbf{Q} \right)  \nonumber\\
  &  & \times f_{ij} \left( \tmop{rel} \left( \mathbf{\tmmathbf{p}}_{\perp
  \tmmathbf{Q}}, \tmmathbf{P}_{\perp \tmmathbf{Q}} \right) -
  \frac{\tmmathbf{Q}}{2} + \frac{\mathcal{E}_{ij}}{Q^2 / m_{\ast}}
  \tmmathbf{Q}, \tmop{rel} \left( \mathbf{\tmmathbf{p}}_{\perp \tmmathbf{Q}},
  \tmmathbf{P}_{\perp \tmmathbf{Q}} \right) + \frac{\tmmathbf{Q}}{2} +
  \frac{\mathcal{E}_{ij}}{Q^2 / m_{\ast}} \tmmathbf{Q} \right) \nonumber\\
  &  & \times f^{\ast}_{kl} \left( \tmop{rel} \left(
  \mathbf{\tmmathbf{p}}_{\perp \tmmathbf{Q}}, \tmmathbf{P}_{\perp
  \tmmathbf{Q}} \right) - \frac{\tmmathbf{Q}}{2} + \frac{\mathcal{E}_{kl}}{Q^2
  / m_{\ast}} \tmmathbf{Q}, \tmop{rel} \left( \mathbf{\tmmathbf{p}}_{\perp
  \tmmathbf{Q}}, \tmmathbf{P}_{\perp \tmmathbf{Q}} \right) +
  \frac{\tmmathbf{Q}}{2} + \frac{\mathcal{E}_{kl}}{Q^2 / m_{\ast}}
  \tmmathbf{Q} \right) \nonumber\\
  &  & \times \left\{ E^{^{\dag}}_{kl} E_{ij}, \rho (\tmmathbf{P}) \right\} .
  \label{eq:cbbemu}
\end{eqnarray}
Let us now recall the expression of the dynamic structure factor for a gas of
free particles obeying Maxwell-Boltzmann statistics, which is given by
\begin{eqnarray}
  S_{\tmop{MB}} \left( \mathbf{\tmmathbf{Q}}, E \right) & = &
  \sqrt{\frac{\beta m}{2 \pi}} \frac{1}{Q} \exp \left( - \frac{\beta}{8 m}
  \frac{(Q^2 + 2 mE)^2}{Q^2} \right),  \label{eq:mb}
\end{eqnarray}
where the variables $\tmmathbf{Q}$ and $E$ denote energy transfer and momentum
transfer in a scattering event. The dynamic structure factor is a two-point
correlation function appearing in the expression of the scattering cross
section of a probe scattering off a macroscopic sample, in the present case
the gas, expressed in terms of momentum and energy transferred in the
collision {\cite{Schwabl2003,Pitaevskii2003}}. Its general expression is given
by the Fourier transform with respect to energy and momentum transfer of the
density-density correlation function of the medium, and for the case of a
sample of non interacting particles can be analytically evaluated to give Eq.
(\ref{eq:mb}). The physical meaning of the dynamic structure factor for the
characterization of scattering of a test particle off a gas explains its
natural appearance in the expression of the quantum linear Boltzmann equation,
as already recognized in
{\cite{Vacchini2001a,Vacchini2001b,Petruccione2005a,Hornberger2008a}}. As we
are now going to show the dynamic structure factor also appears when
considering internal degrees of freedom, the energy transfer being now also
related to the energy absorbed or released as a consequence of internal
transitions. Exploiting Eq. (\ref{eq:perp}) we observe in fact the identity
\begin{eqnarray}
  \mu_{\beta} \left( \tmmathbf{p} \mathbf{}_{\perp \tmmathbf{Q}} \text{$+
  \frac{m}{m_{\ast}^{}} \frac{\tmmathbf{Q}}{2} + \frac{m}{M} \tmmathbf{P}_{\|
  \tmmathbf{Q}}$} + \frac{\mathcal{E}_{ij}}{Q^2 / m} \tmmathbf{Q} \right) & =
  & \mu_{\beta} \left( \tmmathbf{p} \mathbf{}_{\perp \tmmathbf{Q}} \text{$+
  \left( \frac{Q^2 + 2 m \left( E \left( \tmmathbf{Q}, \tmmathbf{P} \right) +
  \mathcal{E}_{ij} \right)}{Q^2} \right) \frac{\tmmathbf{Q}}{2}$} \right)
  \nonumber
\end{eqnarray}
leading via Eq. (\ref{eq:muMB}) and Eq. (\ref{eq:mb}) to
\begin{eqnarray}
  \frac{m}{Q} \mu_{\beta} \left( \tmmathbf{p} \mathbf{}_{\perp \tmmathbf{Q}}
  \text{$+ \frac{m}{m_{\ast}^{}} \frac{\tmmathbf{Q}}{2} + \frac{m}{M}
  \tmmathbf{P}_{\| \tmmathbf{Q}}$} + \frac{\mathcal{E}_{ij}}{Q^2 / m}
  \tmmathbf{Q} \right) & = & \mu_{\beta} \left( \tmmathbf{p}_{\bot
  \tmmathbf{Q}} \right) S_{MB} \left( \tmmathbf{Q}, E \left( \tmmathbf{Q},
  \tmmathbf{P} \right) + \mathcal{E}_{ij} \right),  \label{eq:equivbis}
\end{eqnarray}
where $\mu_{\beta} \left( \tmmathbf{p}_{\bot \tmmathbf{Q}} \right) $ denotes
the Maxwell-Boltzmann distribution in two dimensions. The quantity
\begin{eqnarray}
  &  & E \left( \tmmathbf{Q}, \tmmathbf{P} \right) = \frac{\left(
  \tmmathbf{P}+\tmmathbf{Q} \right)^2}{2 M} - \frac{P^2}{2 M} = \frac{Q^2}{2
  M} + \frac{\tmmathbf{Q} \cdot \tmmathbf{P}}{M},  \label{eq:etransfer}
\end{eqnarray}
actually only depending on $\tmmathbf{P}_{\|\tmmathbf{Q}}$, is the energy
transferred to the centre of mass in a collision in which the momentum of the
test particle changes from $\tmmathbf{P}$ to $\tmmathbf{P}+\tmmathbf{Q}$.
Relying on Eq. (\ref{eq:equivbis}) and Eq. (\ref{eq:perp}) we can therefore
finally write
\begin{eqnarray}
  \frac{\mathd}{\tmop{dt}} \rho (\tmmathbf{P}) & = &
  \frac{n_{\tmop{gas}}}{m^2_{\ast}} \sum_{\tmscript{\begin{subarray}{c}
    ijkl\\
    \mathcal{E}_{ij} = \mathcal{E}_{kl}
  \end{subarray}}} \bigintlim \mathd \tmmathbf{Q} \int_{\tmmathbf{Q}^{\bot}}
  \bignone \mathd \tmmathbf{p} \hspace{0.2em} \mu_{\beta} \left(
  \tmmathbf{p}_{\bot \tmmathbf{Q}} \right) S_{MB} \left( \tmmathbf{Q}, E
  \left( \tmmathbf{Q}, \tmmathbf{P}-\tmmathbf{Q} \right) + \mathcal{E}_{ij}
  \right) \nonumber\\
  &  & \times f_{ij} \left( \tmop{rel} \left( \mathbf{\tmmathbf{p}}_{\perp
  \tmmathbf{Q}}, \tmmathbf{P}_{\perp \tmmathbf{Q}} \right) -
  \frac{\tmmathbf{Q}}{2} + \frac{\mathcal{E}_{ij}}{Q^2 / m_{\ast}}
  \tmmathbf{Q}, \tmop{rel} \left( \mathbf{\tmmathbf{p}}_{\perp \tmmathbf{Q}},
  \tmmathbf{P}_{\perp \tmmathbf{Q}} \right) + \frac{\tmmathbf{Q}}{2} +
  \frac{\mathcal{E}_{ij}}{Q^2 / m_{\ast}} \tmmathbf{Q} \right) \, \nonumber\\
  &  & \times f^{\ast}_{kl} \left( \tmop{rel} \left(
  \mathbf{\tmmathbf{p}}_{\perp \tmmathbf{Q}}, \tmmathbf{P}_{\perp
  \tmmathbf{Q}} \right) - \frac{\tmmathbf{Q}}{2} + \frac{\mathcal{E}_{kl}}{Q^2
  / m_{\ast}} \tmmathbf{Q}, \tmop{rel} \left( \mathbf{\tmmathbf{p}}_{\perp
  \tmmathbf{Q}}, \tmmathbf{P}_{\perp \tmmathbf{Q}} \right) +
  \frac{\tmmathbf{Q}}{2} + \frac{\mathcal{E}_{kl}}{Q^2 / m_{\ast}}
  \tmmathbf{Q} \right) \nonumber\\
  &  & \times E_{ij} \rho (\tmmathbf{P}-\tmmathbf{Q}) E^{^{\dag}}_{kl}
  \nonumber\\
  &  & - \frac{1}{2} \frac{n_{\tmop{gas}}}{m^2_{\ast}}
  \sum_{\tmscript{\begin{subarray}{c}
    ijkl\\
    \mathcal{E}_{ij} = \mathcal{E}_{kl}
  \end{subarray}}} \bigintlim \mathd \tmmathbf{Q} \int_{\tmmathbf{Q}^{\bot}}
  \bignone \mathd \tmmathbf{p} \hspace{0.2em} \mu_{\beta} \left(
  \tmmathbf{p}_{\bot \tmmathbf{Q}} \right) S_{MB} \left( \tmmathbf{Q}, E
  \left( \tmmathbf{Q}, \tmmathbf{P} \right) + \mathcal{E}_{ij} \right)
  \nonumber\\
  &  & \times f_{ij} \left( \tmop{rel} \left( \mathbf{\tmmathbf{p}}_{\perp
  \tmmathbf{Q}}, \tmmathbf{P}_{\perp \tmmathbf{Q}} \right) -
  \frac{\tmmathbf{Q}}{2} + \frac{\mathcal{E}_{ij}}{Q^2 / m_{\ast}}
  \tmmathbf{Q}, \tmop{rel} \left( \mathbf{\tmmathbf{p}}_{\perp \tmmathbf{Q}},
  \tmmathbf{P}_{\perp \tmmathbf{Q}} \right) + \frac{\tmmathbf{Q}}{2} +
  \frac{\mathcal{E}_{ij}}{Q^2 / m_{\ast}} \tmmathbf{Q} \right) \nonumber\\
  &  & \times f^{\ast}_{kl} \left( \tmop{rel} \left(
  \mathbf{\tmmathbf{p}}_{\perp \tmmathbf{Q}}, \tmmathbf{P}_{\perp
  \tmmathbf{Q}} \right) - \frac{\tmmathbf{Q}}{2} + \frac{\mathcal{E}_{kl}}{Q^2
  / m_{\ast}} \tmmathbf{Q}, \tmop{rel} \left( \mathbf{\tmmathbf{p}}_{\perp
  \tmmathbf{Q}}, \tmmathbf{P}_{\perp \tmmathbf{Q}} \right) +
  \frac{\tmmathbf{Q}}{2} + \frac{\mathcal{E}_{kl}}{Q^2 / m_{\ast}}
  \tmmathbf{Q} \right) \nonumber\\
  &  & \times \left\{ E^{^{\dag}}_{kl} E_{ij}, \rho (\tmmathbf{P}) \right\}, 
  \label{eq:final}
\end{eqnarray}
where because of the delta of energy conservation Eq. (\ref{eq:ec}) exploited
in coming to this final expression the transition energy $\mathcal{E}_{ij}$
must be equal to zero whenever $\tmmathbf{Q}$ is equal to zero, due to the
fact that we consider the gas particles as structureless. We stress the
dependence on $\tmmathbf{P}-\tmmathbf{Q}$ in the gain term with respect to
$\tmmathbf{P}$ in the loss term. Note also the very natural appearance of the
argument $E \left( \tmmathbf{Q}, \tmmathbf{P} \right) + \mathcal{E}_{ij}$ in
the dynamic structure factor, corresponding to the energy transfer in the
interaction events, due to both the momentum exchange and the internal
transition, whenever the scattering is not elastic.

Eq. \eqref{eq:final} is equivalent to Eq. \eqref{eq:clbeMB}, but it is
written in a more convenient way for the sake of considering a quantum
description of the centre of mass degrees of freedom. The quantum
master equation for the dynamics of both internal and centre of mass
degrees of freedom has to be of Lindblad form and to coincide with the
semiclassical expression \eqref{eq:final} when considering the
diagonal matrix elements in the momentum representation. Moreover due
to  the homogeneity of the gas the equation has to reflect the
physical invariance under translations, which is expressed at the
level of the master equation by the property of covariance under
translations, corresponding to the fact that the generator of the
master equation commutes with the generator of translations. This
property has been considered at a formal level in
\cite{Holevo1993a,Holevo1993b,Holevo1995a,Holevo1998}, leading to a
general mathematical characterization of Lindblad structures complying
with translational invariance, and discussed in a physical framework
in \cite{Vacchini2001b,Petruccione2005a}. In view of these
requirements the quantum master equation is simply obtained by making operator-valued the
relevant physical expressions appearing in the equation and depending on the
momentum of the test particle, such as dynamic structure factor and scattering
amplitude. In this way one obtains the following master equation for a statistical
operator $\varrho$ on the space $\mathcal{} L^2 \left( \mathbbm{R}^3 \right)
\otimes \mathbbm{C}^n \text{}$, which is manifestly in Lindblad form
\begin{eqnarray}
  \frac{\mathd}{\tmop{dt}} \varrho & = & \frac{n_{\tmop{gas}}}{m^2_{\ast}}
  \sum_{\tmscript{\begin{subarray}{c}
    ijkl\\
    \mathcal{E}_{ij} = \mathcal{E}_{kl}
  \end{subarray}}} \bigintlim \mathd \tmmathbf{Q} \int_{\tmmathbf{Q}^{\bot}}
  \bignone \mathd \tmmathbf{p} \hspace{0.2em} \mu_{\beta} \left(
  \tmmathbf{p}_{\bot \tmmathbf{Q}} \right)  \nonumber\\
  &  & \left[ f_{ij} \left( \tmop{rel} \left( \mathbf{\tmmathbf{p}}_{\perp
  \tmmathbf{Q}}, \mathsf{P}_{\perp \tmmathbf{Q}} \right) -
  \frac{\tmmathbf{Q}}{2} + \frac{\mathcal{E}_{ij}}{Q^2 / m_{\ast}}
  \tmmathbf{Q}, \tmop{rel} \left( \mathbf{\tmmathbf{p}}_{\perp \tmmathbf{Q}},
  \mathsf{P}_{\perp \tmmathbf{Q}} \right) + \frac{\tmmathbf{Q}}{2} +
  \frac{\mathcal{E}_{ij}}{Q^2 / m_{\ast}} \tmmathbf{Q} \right) \right.
  \nonumber\\
  &  & \times \mathe^{i\tmmathbf{Q} \cdot \mathsf{X} / \hbar} \sqrt{S_{MB}
  \left( \tmmathbf{Q}, E \left( \tmmathbf{Q}, \mathsf{P} \right) +
  \mathcal{E}_{ij} \right)} E_{ij} \varrho E^{^{\dag}}_{kl} \sqrt{S_{MB}
  \left( \tmmathbf{Q}, E \left( \tmmathbf{Q}, \mathsf{P} \right) +
  \mathcal{E}_{kl} \right)} \mathe^{- i\tmmathbf{Q} \cdot \mathsf{X} / \hbar}
  \nonumber\\
  &  & \left. \times f_{kl}^{\dag} \left( \tmop{rel} \left(
  \mathbf{\tmmathbf{p}}_{\perp \tmmathbf{Q}}, \mathsf{P}_{\perp \tmmathbf{Q}}
  \right) - \frac{\tmmathbf{Q}}{2} + \frac{\mathcal{E}_{kl}}{Q^2 / m_{\ast}}
  \tmmathbf{Q}, \tmop{rel} \left( \mathbf{\tmmathbf{p}}_{\perp \tmmathbf{Q}},
  \mathsf{P}_{\perp \tmmathbf{Q}} \right) + \frac{\tmmathbf{Q}}{2} +
  \frac{\mathcal{E}_{kl}}{Q^2 / m_{\ast}} \tmmathbf{Q} \right) \right]
\nonumber\\
&  & - \frac{1}{2} \frac{n_{\tmop{gas}}}{m^2_{\ast}}
  \sum_{\tmscript{\begin{subarray}{c}
    ijkl\\
    \mathcal{E}_{ij} = \mathcal{E}_{kl}
  \end{subarray}}} \bigintlim \mathd \tmmathbf{Q} \int_{\tmmathbf{Q}^{\bot}}
  \bignone \mathd \tmmathbf{p} \hspace{0.2em} \mu_{\beta} \left(
  \tmmathbf{p}_{\bot \tmmathbf{Q}} \right)  \nonumber\\
  &  & \left\{ f_{kl}^{\dag} \left( \tmop{rel} \left(
  \mathbf{\tmmathbf{p}}_{\perp \tmmathbf{Q}}, \mathsf{P}_{\perp \tmmathbf{Q}}
  \right) - \frac{\tmmathbf{Q}}{2} + \frac{\mathcal{E}_{kl}}{Q^2 / m_{\ast}}
  \tmmathbf{Q}, \tmop{rel} \left( \mathbf{\tmmathbf{p}}_{\perp \tmmathbf{Q}},
  \mathsf{P}_{\perp \tmmathbf{Q}} \right) + \frac{\tmmathbf{Q}}{2} +
  \frac{\mathcal{E}_{kl}}{Q^2 / m_{\ast}} \tmmathbf{Q} \right) \right.
  \nonumber\\
  &  & \times f_{ij} \left( \tmop{rel} \left( \mathbf{\tmmathbf{p}}_{\perp
  \tmmathbf{Q}}, \mathsf{P}_{\perp \tmmathbf{Q}} \right) -
  \frac{\tmmathbf{Q}}{2} + \frac{\mathcal{E}_{ij}}{Q^2 / m_{\ast}}
  \tmmathbf{Q}, \tmop{rel} \left( \mathbf{\tmmathbf{p}}_{\perp \tmmathbf{Q}},
  \mathsf{P}_{\perp \tmmathbf{Q}} \right) + \frac{\tmmathbf{Q}}{2} +
  \frac{\mathcal{E}_{ij}}{Q^2 / m_{\ast}} \tmmathbf{Q} \right) \nonumber\\
  &  & \left. \times S_{MB} \left( \tmmathbf{Q}, E \left( \tmmathbf{Q},
  \mathsf{P} \right) + \mathcal{E}_{ij} \right) E^{^{\dag}}_{kl} E_{ij},
  \varrho \right\},  \label{eq:almost}
\end{eqnarray}
where $\mathsf{X}$ and $\mathsf{P}$ denote position and momentum operator of
the massive test particle, and the scattering amplitudes $f_{ij}$ appearing
operator-valued describe inelastic scattering with a momentum transfer
$\tmmathbf{Q}$, between two channels differing in energy by
$\mathcal{E}_{ij}$. One immediately checks that the diagonal matrix elements
in the momentum representation of Eq. (\ref{eq:almost}) do coincide with Eq.
(\ref{eq:final}), and furthermore that neglecting the internal degrees of
freedom one comes back to the quantum linear Boltzmann equation
{\cite{Hornberger2008a}}, which together with the correct behavior under
translations, granted by the very operator structure of the equation, provides
a further argument for the assessment of the off-diagonal matrix
elements. 
The step leading from Eq. \eqref{eq:final} to Eq. \eqref{eq:almost},
which corresponds to promote the classical momentum to the
corresponding operator, similarly to what happens in standard
quantization procedures, relies on the specific structure of
Eq. \eqref{eq:final}, and can also be applied when neglecting the
internal degrees of freedom, in which case it leads to the correct
version of the quantum linear Boltzmann equation, as confirmed by the
independent derivations
\cite{Vacchini2000a,Vacchini2001a,Hornberger2006b,Hornberger2008a}. Of
course the ultimate justification for Eq. \eqref{eq:almost} relies on a
microscopic derivation, which can be obtained similarly but with much
lengthier calculations than in \cite{Hornberger2008a}.
Due to the quite complicated expression it is worth introducing a more
compact notation by defining the Lindblad operators
\begin{eqnarray}
  \mathsf{L} _{\tmmathbf{Q}, \tmmathbf{p}, \mathcal{E}} & = &
  \mathe^{i\tmmathbf{Q} \cdot \mathsf{X} / \hbar} L \left( \tmmathbf{p},
  \mathsf{P} ; \tmmathbf{Q}, \mathcal{E} \right),  \label{eq:l}
\end{eqnarray}
where
\begin{eqnarray}
  L \left( \tmmathbf{p}, \mathsf{P} ; \tmmathbf{Q}, \mathcal{E} \right) & = &
  \sum_{\tmscript{\begin{subarray}{c}
    ij\\
    \mathcal{E}_{ij} = \mathcal{E}_{}
  \end{subarray}}} \bignone f_{ij} \left( \tmop{rel} \left(
  \mathbf{\tmmathbf{p}}_{\perp \tmmathbf{Q}}, \mathsf{P}_{\perp \tmmathbf{Q}}
  \right) - \frac{\tmmathbf{Q}}{2} + \frac{\mathcal{E}_{ij}}{Q^2 / m_{\ast}}
  \tmmathbf{Q}, \tmop{rel} \left( \mathbf{\tmmathbf{p}}_{\perp \tmmathbf{Q}},
  \mathsf{P}_{\perp \tmmathbf{Q}} \right) + \frac{\tmmathbf{Q}}{2} +
  \frac{\mathcal{E}_{ij}}{Q^2 / m_{\ast}} \tmmathbf{Q} \right) \nonumber\\
  &  & \times \sqrt{\frac{n_{\tmop{gas}} }{m_{\ast}^2 } \mu_{\beta} \left(
  \tmmathbf{p}_{\bot \tmmathbf{Q}} \right)} \sqrt{S_{MB} \left( \tmmathbf{Q},
  E \left( \tmmathbf{Q}, \mathsf{P} \right) + \mathcal{E}_{ij} \right)}^{}
  E_{ij},  \label{eq:ll}
\end{eqnarray}
thus writing Eq. (\ref{eq:almost}) in the compact and manifestly Lindblad form
\begin{eqnarray}
  \frac{\mathd}{\tmop{dt}} \varrho & = & \sum_{\mathcal{E}} \bignone
  \bigintlim \mathd \tmmathbf{Q} \int_{\tmmathbf{Q}^{\bot}} \bignone \mathd
  \tmmathbf{p} \left[ \mathsf{L} _{\tmmathbf{Q}, \tmmathbf{p}, \mathcal{E}}
  \varrho \mathsf{L} _{\tmmathbf{Q}, \tmmathbf{p}, \mathcal{E}}^{\dag} -
  \frac{1}{2} \left\{ \mathsf{L} _{\tmmathbf{Q}, \tmmathbf{p},
  \mathcal{E}}^{\dag} \mathsf{L} _{\tmmathbf{Q}, \tmmathbf{p}, \mathcal{E}},
  \varrho \right\} \right] .  \label{eq:qbbe}
\end{eqnarray}
We will refer to Eq. (\ref{eq:qbbe}) or equivalently Eq. (\ref{eq:almost}) as
quantum Bloch-Boltzmann equation.

\section{Reduced non-Markovian dynamics\label{sec:nm}}

We now want to point out two different situations of physical relevance in
which relying on Eq. (\ref{eq:qbbe}) one can obtain a description of
non-Markovian behaviors, typically showing up in non exponential decay e.g. of
coherences of the system under study. Despite focussing on a concrete class of
physical systems, our analysis generally applies to the case of a bipartite
quantum system interacting with a reservoir, provided all degrees of freedom
of the bipartite system are involved in the interaction mechanism between
system and reservoir, thus generating the entanglement which accounts for the
memory effects. This provides a realization and clarification of the scheme
envisaged in {\cite{Budini2005b,Budini2006a}}, calling for extra fictitious
unobserved degrees of freedom in order to lead to a Lindblad rate equation
realizing in the Born approximation a generalized Lindblad structure. In
particular our result goes beyond the Born approximation and displays the
full-fledged generalized Lindblad structure Eq. (\ref{eq:S}), allowing for
truly coupled equations for the different subcollections $\rho_{\alpha}$ \ and
considering both the case of a discrete and a continuous label $\alpha$.

As we discussed in Sect. \ref{sec:intro} the non-Markovian behavior
described via Eq. (\ref{eq:S}) arises when one goes over from Eq.
(\ref{eq:qbbe}) to a semiclassical description, and a classical label
characterizing the initial state cannot be resolved or accounted for
in the final measurement. This semiclassical picture of the dynamics
holds if the initial state of the system is prepared so that one of
the two degrees of freedom is in a classical state, or if decoherence
affects the two kind of degrees of freedom of the bipartite state on
different time scales, so that e.g. the motional dynamics can be
treated classically while the internal degrees of freedom still
require a full quantum treatment.  In this framework knowledge about
the way in which the system is prepared usually naturally provides
information about both parts of the bipartite system, while the final
detection scheme is not necessarily fine enough to fully characterize
the outgoing state. In different contexts it might also possibly arise
as a consequence of superselection rules.

\subsection{Description of centre of mass degrees of freedom\label{sec:cm}}

Let us consider first a situation in which we put our test particle, or
equivalently a sufficiently dilute collection of such test particles so that
they can be considered as non interacting, in a dense inert gas. The test
particle will undergo many collisions quickly leading to a classical
characterization of the motion of its centre of mass, so that only the
diagonal matrix elements in the momentum representation of Eq. (\ref{eq:qbbe})
are left on a time scale set by the collisional decoherence mechanism, which
leads us back to Eq. (\ref{eq:final}), which is also called Bloch-Boltzmann
equation. Of course due to the complexity of the master equation Eq.
(\ref{eq:qbbe}) such a behavior, though naturally expected on physical grounds
and usually invoked in the literature on decoherence {\cite{Joos2003a}},
cannot be easily demonstrated in realistic situations. It has however been
confirmed by means of Monte Carlo simulations, which also allow for estimates
of the decoherence rates {\cite{Breuer2007c}}. It is convenient to write the
equation in a more compact way introducing the following $\mathbbm{C}$-number
rate operators
\begin{eqnarray}
  M^{jl}_{ik} \left( \tmmathbf{P}+\tmmathbf{Q}; \tmmathbf{Q} \right) & = &
  \delta_{\mathcal{E}_{ij}, \mathcal{E}_{kl}}
  \frac{n_{\tmop{gas}}}{m^2_{\ast}} \int_{\tmmathbf{Q}^{\bot}} \bignone \mathd
  \tmmathbf{p} \hspace{0.2em} \mu_{\beta} \left( \tmmathbf{p}_{\bot
  \tmmathbf{Q}} \right) S_{MB} \left( \tmmathbf{Q}, E \left( \tmmathbf{Q},
  \tmmathbf{P} \right) + \mathcal{E}_{ij} \right)  \label{eq:epl}\\
  &  & \times f_{ij} \left( \tmop{rel} \left( \mathbf{\tmmathbf{p}}_{\perp
  \tmmathbf{Q}}, \tmmathbf{P}_{\perp \tmmathbf{Q}} \right) -
  \frac{\tmmathbf{Q}}{2} + \frac{\mathcal{E}_{ij}}{Q^2 / m_{\ast}}
  \tmmathbf{Q}, \tmop{rel} \left( \mathbf{\tmmathbf{p}}_{\perp \tmmathbf{Q}},
  \tmmathbf{P}_{\perp \tmmathbf{Q}} \right) + \frac{\tmmathbf{Q}}{2} +
  \frac{\mathcal{E}_{ij}}{Q^2 / m_{\ast}} \tmmathbf{Q} \right) \nonumber\\
  &  & \times f^{\ast}_{kl} \left( \tmop{rel} \left(
  \mathbf{\tmmathbf{p}}_{\perp \tmmathbf{Q}}, \tmmathbf{P}_{\perp
  \tmmathbf{Q}} \right) - \frac{\tmmathbf{Q}}{2} + \frac{\mathcal{E}_{kl}}{Q^2
  / m_{\ast}} \tmmathbf{Q}, \tmop{rel} \left( \mathbf{\tmmathbf{p}}_{\perp
  \tmmathbf{Q}}, \tmmathbf{P}_{\perp \tmmathbf{Q}} \right) +
  \frac{\tmmathbf{Q}}{2} + \frac{\mathcal{E}_{kl}}{Q^2 / m_{\ast}}
  \tmmathbf{Q} \right), \nonumber
\end{eqnarray}
which provide the rates for scattering from $\tmmathbf{P}$ to
$\tmmathbf{P}+\tmmathbf{Q}$, including the dependence on the indexes for
internal degrees of freedom and the transition energy $\mathcal{E}_{ij}$. In
the limit of an infinitely massive test particle these rate coefficients can
be checked to go over to those derived in {\cite{Hornberger2007b}} for the
case of an immobile system. Exploiting the expression Eq. (\ref{eq:epl}) for
the rate operators we can write Eq. \eqref{eq:final} more compactly as
\begin{eqnarray}
  \frac{\mathd}{\tmop{dt}} \rho (\tmmathbf{P}) =
  \sum_{\tmscript{\begin{subarray}{c}
    ijkl
  \end{subarray}}} \bigintlim \mathd \tmmathbf{Q} \left[ M^{jl}_{ik} \left(
  \tmmathbf{P}; \tmmathbf{Q} \right) E_{ij} \rho (\tmmathbf{P}-\tmmathbf{Q})
  E^{^{\dag}}_{kl} - \frac{1}{2} M^{jl}_{ik} \left( \tmmathbf{P}+\tmmathbf{Q};
  \tmmathbf{Q} \right) \left\{ E^{^{\dag}}_{kl} E_{ij}, \rho (\tmmathbf{P})
  \right\} \right] .  \label{eq:rate} &  & 
\end{eqnarray}

Despite the enormous complexity of this integro-differential operator equation
one can consider some simplified situations, allowing to put into evidence
non-Markovian behaviors arising for the reduced statistical operator $\rho =
\int \mathd \tmmathbf{P} \rho (\tmmathbf{P}) \bignone$, only describing the
internal degrees of freedom, once the dynamics of the various subcollections
$\left\{ \text{$\rho \left( \tmmathbf{P} \right)$} \right\}_{\tmmathbf{P} \in
\mathbbm{R}^3}$ is given by Eq. (\ref{eq:rate}). Let us consider the most
simple conceivable situation, taking an internal $\mathbbm{C}^2$ space and
only allowing for elastic scattering. We thus have $M^{jl}_{ik} \propto
\delta_{ij} \delta_{kl}$, and further restricting to forward scattering we can
write
\begin{eqnarray}
  M^{jl}_{ik} \left( \tmmathbf{P}+\tmmathbf{Q}; \tmmathbf{Q} \right) & = &
  \delta_{ij} \delta_{kl} \delta^3 \left( \tmmathbf{Q} \right) \xi_{ik} \left(
  \tmmathbf{P} \right), \nonumber
\end{eqnarray}
parametrizing the rate operators by means of the functions $\xi_{ik} \left(
\tmmathbf{P} \right)$. According to Eq. (\ref{eq:base}) we denote by $E_{ii} =
|i \rangle \langle i|$ the maps between the same energy eigenstates,
corresponding to the projectors on the two one-dimensional subspaces of
$\mathbbm{C}^2$, so that Eq. (\ref{eq:rate}) now simplifies to
\begin{eqnarray}
  \frac{\mathd}{\tmop{dt}} \rho (\tmmathbf{P}) & = &
  \sum_{\tmscript{\begin{subarray}{c}
    ik
  \end{subarray}}} \xi_{ik} \left( \tmmathbf{P} \right) E_{ii} \rho
  (\tmmathbf{P}) E_{kk} - \frac{1}{2} \left\{ \sum_{\tmscript{\begin{subarray}{c}
    i
  \end{subarray}}} \xi_{ii} \left( \tmmathbf{P} \right) E_{ii}, \rho
  (\tmmathbf{P}) \right\} .  \label{eq:rateeasy}
\end{eqnarray}
The dynamics of the single subcollections $\rho (\tmmathbf{P})$, can now be
easily studied. Setting
\begin{eqnarray}
  \text{$\rho_{ij} (\tmmathbf{P})$} & = & \langle i| \text{$\rho
  (\tmmathbf{P}) |j \rangle$} \nonumber
\end{eqnarray}
for the matrix elements of the collection $\left\{ \text{$\rho \left(
\tmmathbf{P} \right)$} \right\}_{\tmmathbf{P} \in \mathbbm{R}^3}$ of matrices
in $\mathbbm{C}^2$, one immediately sees that there is no dynamics for the
populations, in that $\dot{\rho}_{ii} (\tmmathbf{P}) = 0$, for $i = 1, 2$, so
that in particular integrating over the possible momentum dependence also
$\dot{\rho}_{ii} = 0$, for $i = 1, 2$. The coherences of the subcollections
$\rho_{12} (\tmmathbf{P}) = \rho_{21}^{\ast} (\tmmathbf{P})$ are instead
described by the equation
\begin{eqnarray}
  \frac{\mathd}{\tmop{dt}} \rho_{12} (\tmmathbf{P}) & = & \left[ - \frac{1}{2}
  \xi_{11} \left( \tmmathbf{P} \right) - \frac{1}{2} \xi_{22} \left(
  \tmmathbf{P} \right) + \xi_{12} \left( \tmmathbf{P} \right) \right]
  \rho_{12} (\tmmathbf{P}),  \label{eq:2}
\end{eqnarray}
which introducing what we might call, in the absence of better names, a
momentum dependent friction coefficient
\begin{eqnarray}
  \Xi \left( \tmmathbf{P} \right) & = & \frac{1}{2} \xi_{11} \left(
  \tmmathbf{P} \right) + \frac{1}{2} \xi_{22} \left( \tmmathbf{P} \right) -
  \xi_{12} \left( \tmmathbf{P} \right),  \label{eq:adhoc}
\end{eqnarray}
which in view of Eq. (\ref{eq:epl}) has a positive real part proportional to
the averaged modulus of a difference of forward scattering amplitudes
{\cite{Hornberger2007b}}, is easily solved by
\begin{eqnarray}
  \rho_{12} (\tmmathbf{P}, t) & = & \mathe^{- \Xi \left( \tmmathbf{P} \right)
  t} \rho_{12} (\tmmathbf{P}, 0) . \nonumber
\end{eqnarray}
Let us now consider an initial state of the form $\rho_{12} (\tmmathbf{P}, 0)
= \rho_{12} (0) \mu_{\beta} \left( \tmmathbf{P} \right)$, corresponding to a
preparation in which the test particle, or equivalently the dilute ensemble of
non interacting test particles, is in a classical thermal state as far as
centre of mass is concerned, and has a non vanishing initial value for the
coherences of the internal degrees of freedom. The behavior in time of the
off-diagonal matrix elements observed for the internal degrees of freedom
only, not resolving the momentum of the considered test particle, is then
given by
\begin{eqnarray}
  \rho_{12} (t) & = & \Lambda \left( t \right) \rho_{12} (0), \nonumber
\end{eqnarray}
with
\begin{eqnarray}
  \Lambda \left( t \right) & = & \int \mathd \tmmathbf{P} \bignone \mathe^{-
  \Xi \left( \tmmathbf{P} \right) t} \mu_{\beta} \left( \tmmathbf{P} \right) .
  \label{eq:decay}
\end{eqnarray}
It is now immediately evident that behaviors utterly different from the usual
Markovian exponential decay in time appear depending on the actual expression
of the momentum dependent \ $\Xi \left( \tmmathbf{P} \right)$, the Markovian
case obviously corresponding to a constant friction coefficient $\Xi \left(
\tmmathbf{P} \right) = \eta$.

For the simple case $\Xi \left( \tmmathbf{P} \right) = a\tmmathbf{P}^2$ one
immediately obtains a power law decay of the form
\begin{eqnarray}
  \Lambda \left( t \right) & = & \frac{1}{\left( 1 + t / \tau \right)^{3 /
  2}},  \label{eq:poli}
\end{eqnarray}
where we have set $\tau = 1 / aP_{\beta}^2$, indicating a natural reference
time. Another simple expression of the friction coefficient leads instead to a
stretched exponential. Considering in fact $\Xi \left( \tmmathbf{P} \right) =
b /\tmmathbf{P}^2$ one has to evaluate
\begin{eqnarray}
  \Lambda \left( t \right) & = & \frac{1}{\pi^{3 / 2} P_{\beta}^3} \int \mathd
  \tmmathbf{P} \bignone \mathe^{- \Xi \left( \tmmathbf{P} \right) t}
  \mathe^{-\tmmathbf{P}^2 / P_{\beta}^2},  \label{eq:3}
\end{eqnarray}
so that exploiting the result {\cite{Gradshteyn1965a}}
\begin{eqnarray}
  \int^{\infty}_0 \mathd x \hspace{0.2em} x^2 \mathe^{- a / x^2} \mathe^{-
  bx^2} \bignone & = & \sqrt{\frac{\pi}{16 b^3}} \left( 1 + 2 \sqrt{ab}
  \right) \mathe^{- 2 \sqrt{ab}} \nonumber
\end{eqnarray}
we obtain
\begin{eqnarray}
  \Lambda \left( t \right) & = & \left[ 1 + \left( t / \tau \right)^{1 / 2}
  \right] \mathe^{- \left( t / \tau \right)^{1 / 2}},  \label{eq:stretch}
\end{eqnarray}
describing a stretched exponential decay in time with a square root
correction, where the reference time is now set by $\tau = P_{\beta}^2 /
\left( 4 b \right)$. These two simple choices for the friction coefficient
$\Xi \left( \tmmathbf{P} \right)$, amenable to an analytical treatment, have
clearly shown the appearance of strongly non-Markovian behaviors for the
operator $\rho \left( t \right)$. The considered example is obviously quite
simplified and does not describe in a realistic way all possible aspects of
the dynamics, e.g. the redistribution of population in the internal degrees of
freedom. It allows however to easily grasp some non-Markovian aspects of the
generalized Lindblad structure given by Eq. (\ref{eq:S}), of which Eq.
(\ref{eq:rate}) provides a simple example, even though with a continuous
index.

\subsection{Description of internal degrees of freedom\label{sec:int}}

We now focus on a quite different situation, in which we study the dynamics of
our test particle when flying through an interferometer for massive particles,
e.g. of the Mach-Zender type, as recently realized also for the quantitative
study of decoherence {\cite{Hornberger2003a}}. In such a case the initial
preparation is engineered so as to ensure a coherent superposition of states
of the motional degrees of freedom of the system while, in the absence of a
further selection in the prepared state, the internal degrees of freedom can
be described by a classical distribution, corresponding to a partially
diagonal statistical operator. The diagonal matrix elements of Eq.
(\ref{eq:qbbe}) with respect to the internal degrees of freedom naturally lead
to coupled master equations for the collection $\left\{ \rho_r \right\}_r$ of
trace class operators in $\text{$L^2 \left( \mathbbm{R}^3 \right)$}$, defined
according to $\rho_r = \langle r| \varrho |r \rangle$. In order to keep a
compact notation we introduce the rate operators
\begin{eqnarray}
  \mathsf{R}^{rj} \left( \tmmathbf{p}, \tmmathbf{Q} \right) & = & f_{rj}
  \left( \tmop{rel} \left( \mathbf{\tmmathbf{p}}_{\perp \tmmathbf{Q}},
  \mathsf{P}_{\perp \tmmathbf{Q}} \right) - \frac{\tmmathbf{Q}}{2} +
  \frac{\mathcal{E}_{rj}}{Q^2 / m_{\ast}} \tmmathbf{Q}, \tmop{rel} \left(
  \mathbf{\tmmathbf{p}}_{\perp \tmmathbf{Q}}, \mathsf{P}_{\perp \tmmathbf{Q}}
  \right) + \frac{\tmmathbf{Q}}{2} + \frac{\mathcal{E}_{rj}}{Q^2 / m_{\ast}}
  \tmmathbf{Q} \right) \nonumber\\
  &  & \times \sqrt{\frac{n_{\tmop{gas}} }{m_{\ast}^2 } \mu_{\beta} \left(
  \tmmathbf{p}_{\bot \tmmathbf{Q}} \right)} \mathe^{i\tmmathbf{Q} \cdot
  \mathsf{X} / \hbar} \sqrt{S_{MB} \left( \tmmathbf{Q}, E \left( \tmmathbf{Q},
  \mathsf{P} \right) + \mathcal{E}_{rj} \right)},  \label{eq:ro}
\end{eqnarray}
operator-valued on $\text{$L^2 \left( \mathbbm{R}^3 \right)$}$, so that the
master equations for the subcollections $\left\{ \rho_r \right\}_r$ explicitly
exhibit the generalized Lindblad structure given by Eq. (\ref{eq:S}), with the
additional appearance of integrals over continuous indexes
\begin{eqnarray}
  \frac{\mathd}{\tmop{dt}} \rho_r = \sum_j^{} \bigintlim \mathd \tmmathbf{Q}
  \int_{\tmmathbf{Q}^{\bot}} \bignone \mathd \tmmathbf{p} \left[
  \mathsf{R}^{rj} \left( \tmmathbf{p}, \tmmathbf{Q} \right) \rho_j
  \mathsf{R}^{rj} \left( \tmmathbf{p}, \tmmathbf{Q} \right)^{\dag} -
  \frac{1}{2} \left\{ \mathsf{R}^{jr} \left( \tmmathbf{p}, \tmmathbf{Q}
  \right)^{\dag} \mathsf{R}^{jr} \left( \tmmathbf{p}, \tmmathbf{Q} \right),
  \rho_r \right\} \right] . &  &  \label{eq:rirate}
\end{eqnarray}
The set of master equations given by Eq. (\ref{eq:rirate}) do provide a
non-Markovian dynamics for the statistical operator $\rho = \sum_{r = 1}^n
\rho_r \bignone$ observed at the outcome of the experiment, e.g. to determine
the visibility of the interference fringes, when the detection scheme cannot
resolve the state of the internal degrees of freedom.

In a typical experimental situation for the study of collisional decoherence
one can safely neglect the dependence on the momentum operator $\mathsf{P}$ in
the rate operators defined by Eq. (\ref{eq:ro}), replacing it by the classical
value of the momentum of the incoming test particle, due the fact that on the
decoherence time scale the dissipative dynamics of the momentum does not play
a role {\cite{Vacchini2004a,Hornberger2004a,Hornberger2008a}}. This brings in
an important simplification in Eq. (\ref{eq:rirate}), which can now be written
by putting into evidence the unitary operators $\mathe^{i\tmmathbf{Q} \cdot
\mathsf{X} / \hbar}$, describing the momentum kicks causing decoherence of the
centre of mass, and $\mathbbm{C}$-number positive collision rates
$\lambda_{jr} \left( \tmmathbf{Q} \right)$, which depend on the internal state
of the test particle, thus obtaining
\begin{eqnarray}
  \frac{\mathd}{\tmop{dt}} \rho_r & = & \sum_j^{} \bigintlim \mathd
  \tmmathbf{Q} \left[ \lambda_{rj} \left( \tmmathbf{Q} \right)
  \mathe^{i\tmmathbf{Q} \cdot \mathsf{X} / \hbar} \rho_j \mathe^{-
  i\tmmathbf{Q} \cdot \mathsf{X} / \hbar} - \lambda_{jr} \left( \tmmathbf{Q}
  \right) \rho_r \right] .  \label{eq:1}
\end{eqnarray}
For the case in which the collisions do not lead to transitions between
different internal states, so that $\lambda_{jr} \left( \tmmathbf{Q} \right) =
\delta_{jr} \lambda_r \left( \tmmathbf{Q} \right)$, one comes to the following
master equation describing a dynamics determined by momentum kicks of amount
$\tmmathbf{Q}$, taking place with a probability density $\mathcal{P}_r \left(
\tmmathbf{Q} \right)$ which depends on the internal state of the test particle
\begin{eqnarray}
  \frac{\mathd}{\tmop{dt}} \rho_r & = & \Lambda_r \bigintlim \mathd
  \tmmathbf{Q} \mathcal{P}_r \left( \tmmathbf{Q} \right) \left[
  \mathe^{i\tmmathbf{Q} \cdot \mathsf{X} / \hbar} \rho_r \mathe^{-
  i\tmmathbf{Q} \cdot \mathsf{X} / \hbar} - \rho_r \right] .  \label{eq:decoh}
\end{eqnarray}
In Eq. (\ref{eq:decoh}) the probability density $\mathcal{P}_r \left(
\tmmathbf{Q} \right)$ is defined according to
\begin{eqnarray}
  \text{ $\mathcal{P}_r \left( \tmmathbf{Q} \right)$} & = & \frac{\lambda_r
  \left( \tmmathbf{Q} \right)}{\Lambda_r},  \label{eq:pd}
\end{eqnarray}
with $\Lambda_r = \int \mathd \tmmathbf{Q} \lambda_r \left( \tmmathbf{Q}
\right) \bignone$ the total scattering rate for particles with internal state
$r$. The master equations Eq. (\ref{eq:decoh}) are easily solved in the
position representation according to
\begin{eqnarray}
  \langle \tmmathbf{x}| \rho_r \left( t \right) |\tmmathbf{y} \rangle & = &
  \mathe^{- \Lambda_r \left[ 1 - \phi_r \left( \tmmathbf{x}-\tmmathbf{y}
  \right) \right] t} \langle \tmmathbf{x}| \rho_r \left( 0 \right)
  |\tmmathbf{y} \rangle,  \label{eq:ps}
\end{eqnarray}
with
\begin{eqnarray}
  \phi_r \left( \tmmathbf{x}-\tmmathbf{y} \right) & = & \bigintlim \mathd
  \tmmathbf{Q} \mathcal{P}_r \left( \tmmathbf{Q} \right) \mathe^{i\tmmathbf{Q}
  \cdot \mathsf{\left( \tmmathbf{x}-\tmmathbf{y} \right)} / \hbar}, \nonumber
\end{eqnarray}
the Fourier transform of the probability density given by Eq. (\ref{eq:pd}),
that is to say its characteristic function {\cite{Feller1971}}. In particular
the prefactor is the characteristic function of a compound Poisson process,
composed according to the probability density $\mathcal{P}_r \left(
\tmmathbf{Q} \right)$. This equation describes a quite general physical
situation in which one has a sequence of interaction events between system and
environment, distributed in time according to a Poisson distribution, each one
characterized by a random momentum transfer, drawn according to a certain
probability density fixed by the microphysical interaction mechanism
{\cite{Vacchini2005a,Vacchini2007e}}. At variance with a simple Poisson
process the momentum transfer is not deterministically fixed to be the same in
each collision, but is a random variable depending on the details of the
collision.

We now look at the dynamics of the matrix elements of the whole statistical
operator $\rho = \sum_r^{} \rho_r \bignone$, responsible for the description
of the measurement outcomes at the output of the interferometer. To do this we
consider an initial state of the form $\rho_r \left( 0 \right) = p_r \rho
\left( 0 \right)$, with $p_r \geqslant 0$ and $\sum_r^{} p_r = 1 \bignone$,
corresponding to a preparation in which the populations in the internal states
are distributed according to certain weights $p_r$, while the statistical
operator $\rho \left( 0 \right)$ in $L^2 \left( \mathbbm{R}^3 \right)$
characterizes the quantum state of the centre of mass. We thus come to the
following expression for the solution of the non-Markovian set of master
equations given by Eq. (\ref{eq:decoh})
\begin{eqnarray}
  \langle \tmmathbf{x}| \rho \left( t \right) |\tmmathbf{y} \rangle & = &
  \sum_r^{} p_r \bignone \mathe^{- \Lambda_r \left[ 1 - \phi_r \left(
  \tmmathbf{x}-\tmmathbf{y} \right) \right] t} \langle \tmmathbf{x}| \rho_{}
  \left( 0 \right) |\tmmathbf{y} \rangle,  \label{eq:decade}
\end{eqnarray}
where the multiplicative prefactor determines the weight of the matrix
elements in the position representation, both with elapsing time, and as a
function of the distance $\left| \tmmathbf{x}-\tmmathbf{y} \right|$. Such an
expression provides information on the loss of coherence responsible for
reduction in the visibility of the interference fringes. If only one of the
weights $p_r$ is different from zero, and therefore equal to one, one falls
back to the usual Markovian exponential decay in time, possibly with a
modulation in the spatial dependence. \ In particular the coherence of the
quantum state over spatially separated points depends on the details of the
functions $\phi_r \left( \tmmathbf{x}-\tmmathbf{y} \right)$, given by the
Fourier transform of the probability density of momentum kicks in the
scattering events. In a typical situation such functions quickly go to zero in
their dependence on the distance $\left| \text{$\tmmathbf{x}-\tmmathbf{y}$}
\right|$, so that e.g. in the Markovian case one is simply left with a
constant exponential loss of visibility {\cite{Hornberger2003a}}. With the
present more general initial state even for $\phi_r \left(
\tmmathbf{x}-\tmmathbf{y} \right) \simeq 0$ one has a non trivial structure
describing a decay of coherence other than exponential
\begin{eqnarray}
  \langle \tmmathbf{x}| \rho \left( t \right) |\tmmathbf{y} \rangle & \simeq &
  \Psi \left( t \right) \langle \tmmathbf{x}| \rho_{} \left( 0 \right)
  |\tmmathbf{y} \rangle,  \label{eq:nm}
\end{eqnarray}
where the function
\begin{eqnarray}
  \Psi \left( t \right) & = & \sum_r^{} p_r \bignone \mathe^{- \Lambda_r t}, 
  \label{eq:multi}
\end{eqnarray}
is the survival probability of a multiexponential distribution, i.e. the
probability to have no event up to a time $t$ for such a distribution
{\cite{Feller1971}}. Depending on the weights $p_r$ and the rates $\Lambda_r$,
not only simple deviations from the exponential law can appear, but also
utterly different behaviors. To clarify this point let us consider for the
sake of example a geometric distribution of weights, with ratio $p_0 =
\mathe^{- a}$, $a \in \mathbbm{R_+}$, so that
\begin{eqnarray}
  p_r & = & \left( 1 - p_0 \right) p_0^r, \nonumber
\end{eqnarray}
and a geometric progression of rates
\begin{eqnarray}
  \Lambda_r & = & \Lambda_0 \gamma^r_0, \nonumber
\end{eqnarray}
with ratio $\gamma_0 = \mathe^{- b}$, $b \in \mathbbm{R_+}$, and the reference
rate $\Lambda_0$ as scale factor. The survival probability then reads
\begin{eqnarray}
  \Psi \left( t \right) & = & \left( 1 - p_0 \right) \sum_r^{} p_0^r \mathe^{-
  \gamma^r_0 \Lambda_0 t}, \nonumber
\end{eqnarray}
which due to the relation {\cite{Alemany1997a}}
\begin{eqnarray}
  \Psi \left( \Lambda_0 t \right) & = & \frac{1}{p_0} \left[ \Psi \left( t
  \right) - \left( 1 - p_0 \right) \mathe^{- \Lambda_0 t} \right] \nonumber
\end{eqnarray}
exhibits at long times a power law decay
\begin{eqnarray}
  \Psi \left( \Lambda_0 t \right) & \begin{subarray}{c}
    _{t \gg 1}\\
    \simeq
  \end{subarray} & \frac{1}{\left( \Lambda_0 t \right)^{a / b}} .  \label{eq:pl}
\end{eqnarray}
This \ most simple example allowing for an analytical treatment already shows
the rich variety of non-Markovian behaviors which might arise when one uses
the generalized Lindblad structure given by Eq. (\ref{eq:rirate}), which
provides a further example of the general result Eq. (\ref{eq:S}) for the case
of a sum over a discrete index. In particular Eq. (\ref{eq:rirate}) describes
how the dependence of the scattering events on the internal structure of the
test particle affects the loss of coherence in position space, which in turn
determines the reduction of visibility in an interferometric experiment. As it
appears from Eq. (\ref{eq:pl}) this can lead to very strong deviations from
the exponential decay, such as power law behaviors.

\section{Conclusions\label{sec:ceo}}

In this paper we have considered a class of non-Markovian behaviors, arising
when dealing with a bipartite quantum system interacting with a reservoir. The
concrete bipartite system considered was given by a massive test particle, for
which both internal and centre of mass degrees of freedom have been taken into
account. The reservoir was assumed as a structureless gas, affecting our
test particle through collisions whose microscopic characterization depends on
both its motional and internal state. As a starting point we have derived in
Sect. \ref{sec:qbbe} a new quantum master equation describing such dynamics in
a non perturbative way, expressed by Eq. (\ref{eq:qbbe}), which can also be
termed quantum Bloch-Boltzmann equation in that it describes at the quantum
level both kinds of degrees of freedom. When due to decoherence or features of
the initial preparation one of the two degrees of freedom is to be described
classically, one obtains from the quantum Bloch-Boltzmann equation two
examples of a generalized Lindblad structure recently considered for the
description of non-Markovian dynamics {\cite{Breuer2006a,Breuer2007a}}. Such a
generalized Lindblad expression has been outlined in Sect. \ref{sec:gl},
clarifying its mathematical structure and physical motivation. For the case at
hand non-Markovian effects, leading to decay behaviors of coherences of the
system given by stretched exponentials or power laws instead of simple
exponentials, appear when the degrees of freedom allowing for a classical
description are not resolved in the final measurement, only focussing on the
quantum degrees of freedom. This provides a concrete realization of a proposed
mechanism for the appearance of such generalized Lindblad structures
{\cite{Budini2005b,Budini2006a}}, further clarifying the origin of the
non-Markovian behaviors. These behaviors have been spelled out in Sect.
\ref{sec:nm}, focussing on the dynamics of the internal state of a test
particle interacting with an inert gas, as well as on loss of coherence of a
massive particle flying through an interferometer where it interacts with a
background gas. It is to be stressed that in the physical examples
considered in Sec. \ref{sec:cm} and Sec. \ref{sec:int} one has to deal
with decoupled subcollections of statistical operators, so that the
non-Markovian features arise from the average over the classical index
in the initial condition. The generalized Lindblad structure given by
Eq. \eqref{eq:S} also allows to consider coupled equations for the
different subcollections, and for these situations one naturally
expects a much more complicated non-Markovian dynamics.

It immediately appears that the outlined scheme leading to a class of
non-Markovian evolutions generally applies in the presence of the interaction
of a bipartite quantum system with a quantum environment, when one of the
quantum labels of the system becomes classical and can be averaged over. More
generally such a class of non-Markovian evolutions appear in the presence of a
classical degree of freedom, described by means of some discrete or continuous
label, which is involved in the characterization of the interaction between
two quantum degrees of freedom, and is averaged over in order to give the
relevant dynamics. This classical label might as well appear on the side of
the environment, corresponding to so called structured reservoirs, or on the
side of the system, as in the case of a bipartite system. The present work can
naturally be extended to include an internal structure in the gas particles,
which could also influence the scattering amplitude, introducing new channels.
In particular a detailed analysis of the rate operators based on microphysical
informations could pave the way to new interferometric experiments for the
quantitative study of decoherence, exhibiting more general behaviors than
exponential decay of visibility with elapsed interaction time.

\section{Acknowledgments}

The author is grateful to Heinz-Peter Breuer, Klaus Hornberger and Ludovico
Lanz for helpful discussions and reading of the manuscript.

\end{document}